%Paper: astro-ph/9511095
%From: jijina@pablo.physics.LSA.UMich.EDU (Jasmin Jijina)
%Date: Mon, 20 Nov 1995 17:54:57 GMT

\magnification=1200
\hoffset=0.1truein
\voffset=0.1truein
\vsize=23.0truecm
\hsize=16.25truecm
\parskip=0.2truecm
\def\newpage{\vfill\eject}

\def\pp{\parshape 2 0.0truecm 16.25truecm 2truecm 14.25truecm}

\def\rhobar{ {\langle \rho \rangle}}
\def\muck{ {\mu_C}}
%\baselineskip=20pt
%
%---------------------------------------------------------------------------
%
\vskip 0.5truein
\centerline{\bf INFALL COLLAPSE SOLUTIONS IN THE INNER LIMIT:}
\centerline{\bf RADIATION PRESSURE AND ITS EFFECTS ON STAR FORMATION}
\vskip 0.2truein
\centerline{\bf Jasmin Jijina and Fred C. Adams}
\vskip 0.15truein
\centerline{\it Department of Physics}
\centerline{\it University of Michigan, Ann Arbor, MI 48109}
\vskip 0.15truein
\centerline{and}
\vskip 0.15truein
\centerline{\it Institute for Theoretical Physics}
\centerline{\it University of California, Santa Barbara, CA 93106}
\vskip 0.3truein
\centerline{\it submitted to The Astrophysical Journal: 18 July 1995}
\vskip 0.3truein
\centerline{\it revised: 1 November 1995}
\bigskip

\bigskip
\centerline {\bf Abstract}

In this paper, we study infall collapse solutions for star formation
in the small radius limit where the particle orbits become nearly
pressure-free.  We generalize previous solutions to simultaneously
include the effects of both radiation pressure and angular momentum.
The effects of radiation pressure can be modeled using a modified
potential; for representative cases of such potentials, we obtain
analytical solutions for the density and velocity fields.  In general,
radiation pressure limits the maximum mass of a forming star by
reversing the infall when the star becomes sufficiently large.  Our
results imply that this maximum mass scale is given by the condition
that the turnaround radius $R_R$ (the radius at which the radiation
pressure force exceeds the gravitational force) exceeds the centrifugal
radius $R_C$ (the angular momentum barrier).  The maximum mass scale
for a star forming within a rotating collapse flow with radiation
pressure depends on the initial conditions, but is generally much
larger than for the case of spherical infall considered previously.
In particular, stars with masses $M_\ast$ $\sim 100$ $M_\odot$ can
form for a fairly wide range of initial conditions.

\vskip 0.4truein
\noindent
{\it Subject headings:} stars: formation -- hydrodynamics
\vskip 0.4truein

\newpage
\bigskip
\centerline{\bf 1. INTRODUCTION}
\medskip

In this paper, we study generalized aspects of the collapse of
molecular cloud cores to form stars.  In particular, we include the
effects of radiation pressure for the inner regime of the collapse
flow.  This work thus helps extend the current theory of star
formation to include the formation of stars of higher mass.

In the current theory of star formation, molecular cloud cores
provide the initial conditions for the star formation process.
Star formation occurs when these cores collapse. A pressure
supported star forms at the center of the collapse flow and
a circumstellar disk forms around it from the infalling material
with higher specific angular momentum (cf. the review of Shu,
Adams, \& Lizano 1987).   Thus, during this formative stage,
a protostar consists of a central star/disk system which is
deeply embedded within an infalling envelope of dust and gas.
The characteristics of this infalling envelope largely determine
the manner in which the protostar evolves.

Previous collapse calculations have focused on isothermal cloud cores
which are either perfectly spherical (Larson 1969ab; Shu 1977; Hunter
1977) or slowly rotating (Cassen \& Moosman 1981; Terebey, Shu, \&
Cassen 1984). These solutions have been extremely useful in studies of
low-mass protostars.  In particular, these infall solutions have been
used as a starting point for radiative transfer calculations to
determine the spectral energy distributions of protostellar objects;
the results are in good agreement with observed protostellar
candidates (e.g., Adams, Lada, \& Shu 1987, hereafter ALS; Butner et
al.  1991; Kenyon, Calvet, \& Hartmann 1993). However, with large
amounts of new kinematic data becoming available, more detailed
collapse solutions are desirable.

These previous studies have shown that the collapse flow will remain
nearly spherical in the outer regions, i.e., outside a centrifugal
radius $R_C$ which is determined by conservation of angular momentum.
When magnetic fields are present, an analogous magnetic barrier occurs
at the radius $R_B$ where the Lorentz force of the magnetic field
exceeds that of gravity (see Galli \& Shu 1993ab).  In this present
work, we study the collapse solutions in the inner regime ($r$ $\sim
R_C$), where the infalling particles follow nearly ballistic
trajectories as they spiral into the central star/disk system (Cassen
\& Moosman 1981; Terebey et al. 1984). Previous work in this regime
has been limited to the case of zero energy orbits and the
gravitational potential of a point mass (the star).

In this work, we consider the additional effect of radiation pressure,
which is important for protostellar objects with mass $M \geq 2$
$M_\odot$.  In the current star formation scenario, there is an
``opacity gap'' immediately surrounding the star; here, the dust
evaporates and radiation streams freely through this region.  At the
dust evaporation radius (often denoted as the ``dust destruction
front''), the ultraviolet and visible photons are absorbed in a thin
shell and the photons are thermalized. Outside this radius, the
infalling envelope is optically thick at infrared wavelengths where
the warm dust reradiates. The gradient of this infrared radiation
provides a radiation pressure which acts to decelerate
the infalling gas.

Radiation pressure has been considered previously for spherical
collapse (Larson \& Starrfield 1971; Kahn 1974; Wolfire \& Cassinelli
1986, 1987).  These studies show that radiation pressure severely
restricts the possible masses of forming stars.  Nakano (1989) has
argued that nonspherical infall can lead to the formation of more
massive stars (see also Nakano, Hasegawa, \& Norman 1995); this work
assumes both a flattened density distribution and a nonspherical
radiation field. In this present study, we carry this idea further by
finding a self-consistent infall-collapse solution which includes the
effects of radiation pressure in the inner regime where rotation plays
a large role and the flow is highly non-spherical.  In our work, we
assume a spherical radiation field and then calculate the
corresponding nonspherical density field.  The ``inner'' solutions
resulting from this study can then be matched onto the ``outer
solutions'' obtained elsewhere (e.g., Shu 1977; Adams et al. 1995) to
provide a complete collapse solution for the star formation problem.

This paper is organized as follows.  In \S 2, we formulate the
collapse problem including the effects of radiation pressure
and solve for the infall solution in the inner regime.  In the
next section (\S 3), we find the mass and radial scales in the
problem; in particular, we find the maximum mass of a star/disk
system that can form within this infall scenario. If disk accretion
is efficient, this mass scale is the maximum stellar mass that can
be assembled in the presence of radiation pressure.  We consider several
additional issues in \S 4, including the effects of magnetic fields.
We conclude in \S 5 with a summary and discussion of our results.

\bigskip
\centerline{\bf 2. INFALL WITH RADIATION PRESSURE}
\medskip

In this section, we determine the most important effects of radiation
pressure on the infall collapse solutions. For the cases of interest,
the inner limit of the collapse flow approaches a ballistic
(pressure-free) form.  We expect our solutions to be valid within
the following range of (radial) size scales:
$$R_\ast < r \sim R_C \ll r_H \, . \eqno(2.1)$$
In this ordering constraint, the scale $R_\ast$ is the radius of the
forming star and defines the inner boundary of the collapse flow. The
scale $R_C$ is the centrifugal radius (defined more precisely below)
which very roughly divides the nearly spherical outer region of the
flow from the highly nonspherical inner region. Finally, the scale
$r_H$ is the head of the expansion wave which divides the static outer
core from the collapsing inner core (see Shu 1977).  For radii
comparable to the expansion wave radius $r_H$, pressure effects
cannot be neglected and the solutions found here must be modified.

\bigskip
\centerline{\it 2.1 Coupling of the Radiation Field and the Infalling Gas}
\medskip

The coupling between the radiation field and the infalling material is
determined primarily by the dust opacity $\kappa_\nu$, which specifies
the cross section of the interaction.  We are thus assuming that the
dust and gas are themselves well coupled dynamically. This assumption
should be well satisfied in the dense inner regions of interest here.
In this regime, the radial force $f$ per unit mass exerted on the
infalling gas by the radiation field is given by
$$f = \int_0^\infty {\kappa_\nu L_\nu \over 4 \pi c r^2 } \, d\nu \,
= {L \over 4 \pi c} {\kappa_E (T) \over r^2 } , \eqno(2.2)$$
where $L_\nu = 4 \pi r^2 F_\nu$ and $F_\nu$ is the flux density of
the radiation field.  The second equality essentially defines the
weighted mean opacity $\kappa_E$ and is thus formally exact.
We can ``evaluate'' this force term by approximating the weighted
mean opacity as the Planck mean opacity, i.e.,
$$f \approx {L \over 4 \pi c} {\kappa_P (T) \over r^2}
\approx {L \over 4 \pi c} {\kappa_P (T_0) \over r^2}
{T \over T_0} \, , \eqno(2.3)$$
where $T_0$ is a fiducial temperature.
The use of the Planck mean opacity implicitly
assumes that the protostellar radiation field is thermalized
to the local dust temperature; this condition is satisfied in the
optically thick inner regions of the collapsing envelope where
essentially all of the effects of radiation pressure take place.
In the second approximate equality, we have made the additional
assumption that the Planck mean opacity is a linear function of
temperature; this result is exact for the particular case in which
$\kappa_\nu \propto \nu$, which is a good approximation for the
frequencies (wavelengths) in the near infrared (Draine \& Lee 1984)
where most of the interaction between the radiation field and the
dust takes place.

Notice that we have assumed a spherical radiation field. This
assumption is important because it allows for a completely
analytical treatment of the problem.  For high mass stars,
the stellar component of the radiation field is nearly
spherical and dominates the (nonspherical) radiation field
of the disk.  However, the nonspherical density distribution
and the effects of bipolar outflows make the radiation field
depart from spherical symmetry; these effects are not included
in this present work.

We must determine the temperature distribution for protostellar
envelopes.  To a reasonable degree of approximation, the temperature
profile in protostellar envelopes can be taken to be a power-law over
the range of radial scales of interest.  Simple analytical estimates
show that the power-law index $q$ of the temperature distribution has
the value $q \sim 5/6$ in the optically thick inner regions and flattens
out to the range $q \sim 1/3 - 2/5$ in the optically thin outer
regions (Adams \& Shu 1985).  Thus, as a starting approximation
to span both regimes, we adopt the simple form
$$T (r) = T_0 (r/r_0)^{-1/2} \, . \eqno(2.4)$$
This simple form allows us to obtain analytic solutions which
elucidate the basic physics of this problem.  Once we have this
basic understanding, we can study the effects of different forms
for the temperature distribution.

Putting all of the above results together, we obtain an outward radial
force due to radiation pressure of the form
$$f = {L \over 4 \pi c} {\kappa_P (T_0) \sqrt{r_0} \over r^{5/2}}
\, . \eqno(2.5)$$
Using this form for the radiation pressure force, we can derive an
effective potential.  The total potential (including both radiation
pressure and gravity) can be written in the form
$$V_{\rm eff} = {G M \over r} \Bigl\{ \alpha r^{-1/2} - 1
\Bigr\} \, , \eqno(2.6)$$
where we have defined a parameter $\alpha$,
$$\alpha \equiv {L \kappa_P(T_0) \over 6 \pi G M c} \, \sqrt{r_0}
\, , \eqno(2.7{\rm a})$$
which has units of [length$]^{1/2}$. For the sake of definiteness,
we take the reference radius $r_0$ to be the radius $r_d$ of the
dust destruction front, i.e., the radius at which the temperature
becomes larger than the sublimation temperature of dust grains.
For typical interstellar dust grains, the Planck mean opacity
$\kappa_P$ at this temperature ($T_0 = T_d \sim$ 2300 K) is about
30 cm$^2$/g; for the sake of definiteness, we adopt this value
throughout this paper.  We can thus write the parameter $\alpha$
in the form
$$\alpha \approx 1.6 \times 10^{-3} \, \sqrt{r_d} \, \Bigl[
\, { {\widetilde L} \over {\widetilde M}} \, \Bigr]
\, , \eqno(2.7{\rm b})$$
where we have defined ${\widetilde L} \equiv L/(1 L_\odot)$ and
${\widetilde M} \equiv M/(1 M_\odot)$.  Thus, as expected, for solar
type stars, the correction to the orbits due to radiation pressure
is small.  Radiation pressure begins to play a substantial role in
the infall when $\alpha^2 > {r_d}$; this condition is met for
${\widetilde L}/{\widetilde M} > 640$, which corresponds to
moderately massive stars with $M \sim 7 M_\odot$.

\bigskip
\centerline{\it 2.2 Orbital Solutions}
\nobreak
\medskip

Given the potential, we can find the orbital solutions for parcels of
gas falling in towards the central star.  Our goal is to derive the
functional forms for the velocity fields and the density profile
resulting from the collapse of slowly rotating cloud cores.  This
calculation generalizes the case studied previously (Terebey, Shu, \&
Cassen 1984; Cassen \& Moosman 1981; Ulrich 1976), where the inner
collapse solution is derived using only the gravitational potential
of the star itself.

Since the potential is spherically symmetric, the motion
is confined to a plane and the angular momentum is conserved.
This orbital plane can be described by the coordinates $(r, \phi)$
where the radius $r$ is the same in both the plane and the usual
spherical coordinates. The angle $\phi$ is the angle in the plane
is related to the usual spherical coordinates by the relation
$$\cos \phi = {\cos \theta \over \cos \theta_0 } \, , \eqno(2.8)$$
where $\theta_0$ is the angle of the
asymptotically radial streamline (see below).
We also assume orbits with zero total energy.
For the case of no radiation pressure ($\alpha$ = 0),
the orbits are simply parabolas in the $(r, \phi)$ plane.

For the general case, we can integrate the equations of motion
to obtain the solution for the orbits in the form
$${\mu \over \mu_0} \cos{\widetilde \phi} +
\Bigl[ 1 - {\mu^2 \over \mu_0^2} \Bigr]^{1/2} \sin{\widetilde \phi}
= { 2 [ \zeta r^{1/2} (1 - \mu_0^2) + \alpha ]^2 \over
2 \zeta r (1 - \mu_0^2) + \alpha^2 } - 1 \, . \eqno(2.9)$$
The quantity $\mu_0$ is the cosine of the angle $\theta_0$ of the
asymptotically radial streamline (i.e., the fluid trajectory
that is currently passing through the position given by $\zeta$ and
$\mu \equiv \cos \theta$ initially made the angle $\theta_0$ with
respect to the rotation axis).  The quantity $\zeta$ is defined by
$$\zeta \equiv {j_\infty^2 \over G M r} \,
= {R_C \over r} \, , \eqno(2.10)$$
where $j_\infty$ is the specific angular momentum of parcels of gas
currently arriving at the origin along the equatorial plane.
We have followed previous authors in assuming an initial state
that is rotating at a constant rotation rate $\Omega$, so that
the quantity $j_\infty$ is given by
$$j_\infty = \Omega r_\infty^2 \, , \eqno(2.11)$$
where $r_\infty$ is the starting radius of the material
that is arriving at the origin at a given time.  This radius
can be determined by inverting the initial mass distribution
$M(r)$, which we discuss below. Finally, the angle $\widetilde \phi$
arises from the constant of integration and is defined by
$$\cos {\widetilde \phi} = { \alpha^2 - 2 R_C (1 - \mu_0^2)
\over \alpha^2 + 2 R_C (1 - \mu_0^2) } \, . \eqno(2.12)$$
Using these results, the ``orbit equation'' can be ``simplified''
to the form
$$\Bigl[ \alpha^2 - 2 \zeta r (1-\mu_0^2) \Bigr]
\Bigl[ {\mu \over \mu_0} - 1 \Bigr] +
\Bigl[ 1- {\mu^2 \over \mu_0^2} \Bigr]^{1/2} \alpha
\sqrt{8 \zeta r} (1 - \mu_0^2)^{1/2} =
2 \zeta^2 r (1-\mu_0^2)^2  + 4 \alpha \zeta r^{1/2} (1-\mu_0^2) \, .
\eqno(2.13)$$

The effects of radiation pressure on the infalling orbits can be
illustrated as in Figure 1, which shows projected orbits in the
meridional plane for varying amounts of radiation pressure.
Notice that as the radiation pressure (and hence $\alpha$)
increases, the orbits become increasingly deflected.

When the orbits begin to turn around, adjacent orbits will
intersect each other and the gaseous material will shock.
The locus of this shock surface (or ``turn around surface'')
is shown in Figure 2 for varying amounts of radiation pressure.
Notice that adjacent streamlines do not intersect each other
on the way in.

\bigskip
\centerline{\it 2.3 The Centrifugal Radius}
\medskip

The centrifugal radius $R_C$ plays an important role in
determining the nature of the infall solution.  To evaluate $R_C$,
we must invert the mass distribution of the initial state. Here,
we find $R_C$ for the two types of initial states which are
most applicable for star formation.

We first consider the case of a molecular cloud core described
by an isothermal equation of state.  We thus obtain the profile
$$M(r) = {2 a^2 r \over G} \, \qquad {\rm and \, \, \, \, hence}
\qquad r_\infty  = {GM \over 2 a^2 } \, . \eqno(2.14{\rm a})$$
Thus, for isothermal initial conditions, the centrifugal radius
$R_C$ can be written
$$R_C = {\Omega^2 G^3 M^3 \over 16 a^8} \, , \eqno(2.14{\rm b})$$
where we have combined equations [2.10], [2.11], and [2.14a].

We are also interested in infall solutions which apply to the
formation of more massive stars where isothermal initial conditions
may not apply.  Observations indicate that in regions of molecular
clouds with higher mass clumps, the molecular line widths show a
substantial nonthermal component (e.g., Larson 1985; Myers \& Fuller
1992). Motivated by the finding that these observed molecular
linewidths decrease with density according to the law $\Delta v$
$\propto \rho^{-1/2}$, one can derive a ``logatropic'' equation
of state $P = P_0 \log\rho$ to describe the fluid
(see Lizano \& Shu 1989; Adams et al. 1995).
For this case, the equilibrium mass distribution and
the corresponding $r_\infty$ are given by
$$M(r) = \Bigl[ {2 \pi P_0 \over G} \Bigr]^{1/2} r^2
\, \qquad {\rm and \, \, \, \, hence} \qquad r_\infty  =
M^{1/2} \,  \Bigl[ {2 \pi P_0 \over G} \Bigr]^{-1/4} \, ,
\eqno(2.15{\rm a})$$
where $P_0$ is the pressure scale that determines the amount
of nonthermal support in the cloud. The centrifugal radius for
logatropic initial conditions is given by
$$R_C = {\Omega^2 M \over 2 \pi P_0} \, . \eqno(2.15{\rm b})$$
We note that in the presence of radiation pressure, the effective
centrifugal radius becomes modified, i.e., an effective centrifugal
radius can be defined (see equation [2.29]).

\bigskip
\centerline{\it 2.4 Velocity and Density Fields}
\medskip

Given the orbital solution, we can find the velocity fields
and the corresponding density distribution.  We put together
the geometrical relation [2.8], conservation of angular
momentum, and conservation of energy,
$${1 \over 2} (v_r^2 + v_\theta^2 + v_\varphi^2) =
{GM \over r} - \alpha {G M \over r^{3/2}} \, , \eqno(2.16)$$
to determine the velocity field, which can be written in the form
$$v_r = - \Biggl( {G M \over r} \Biggr)^{1/2}
\Bigl\{ 2 - 2 \alpha r^{-1/2} - \zeta (1 - \mu_0^2)
\Bigr\}^{1/2} , \eqno(2.17)$$
$$v_\theta = \Biggl( {G M \over r} \Biggr)^{1/2}
\Biggl\{ {1 - \mu_0^2 \over 1 - \mu^2 } \, (\mu_0^2 - \mu^2) \,
\zeta \Biggr\}^{1/2} , \eqno(2.18)$$
$$v_\varphi = \Biggl( {G M \over r} \Biggr)^{1/2} \, (1 - \mu_0^2)
\, ( 1 - \mu^2 )^{-1/2} \, \zeta^{1/2} \, . \eqno(2.19)$$
Notice that $\zeta$, $\mu$, and $\mu_0$ are related by the
orbit equation [2.13] so that the velocity field is in fact
completely determined for any position $(r, \theta)$.

The density distribution of the infalling material can be obtained
by applying conservation of mass along a streamtube (Terebey, Shu,
\& Cassen 1984; Chevalier 1983), i.e.,
$$\rho(r,\theta) \, v_r \, r^2
\sin\theta \, d\theta \, d\varphi = - { {\dot M} \over 4 \pi}
\sin\theta_0 \, d\theta_0 \, d\varphi_0 .  \eqno(2.20)$$
In this present context, we ignore the fact that some particle
orbits can turn around and leave the system.  Thus, this solution
for the density is valid for radii larger than the turn around
radius given by equation [3.2] below.  Combining the above equations
allows us to write the density profile in the form
$$\rho(r,\theta) = { {\dot M} \over 4 \pi v_r r^2 }
{d \mu_0 \over d \mu} \, . \eqno(2.21)$$
The properties of the collapsing core
determine the form of ${d\mu_0 / d\mu}$.
In the present case, the orbit equation [2.13] determines
the form of $d\mu_0/d\mu$; thus, with the radial velocity given
by equation [2.17], the density field is completely specified
(analytically, but implicitly).

The density distribution as a function of angle and for varying
amounts of radiation pressure is shown in Figure 3. Two trends
are evident from this figure:  For a given radius, the density
decreases with decreasing angle and with increasing values of
the radiation pressure constant $\alpha$ for small values of
$r/R_C$.

\bigskip
\centerline{\it 2.5 Equivalent Spherical Envelope}
\medskip

One way to characterize the effects of both angular momentum and
radiation pressure is to define an equivalent spherical density
distribution. In other words, we take the angular average of the
nonspherical density distribution (see Adams \& Shu 1986; ALS).
Since one effect of both rotation and radiation pressure is to
prevent material from falling to smaller radii, this equivalent
spherical density distribution will {\it not} be equal to the density
distribution for spherical (nonrotating) initial conditions.  We note
that this approach has proven to be a useful characterization of
protostellar envelopes for purposes of determining their spectral
energy distributions (ALS; Kenyon et al. 1993). The equivalent
spherical density distribution is given by
$$\rhobar = \int_0^{\pi/2} \rho(r, \theta) \sin\theta d\theta \,
= C r^{-3/2} \, {\cal A} (r) \, , \eqno(2.22)$$
where the constant $C$ is defined by
$$C \equiv { {\dot M} \over 4 \pi \sqrt{2 G M} } \, , \eqno(2.23)$$
and where we have defined an asphericity function ${\cal A} (r)$,
$${\cal A} (r) \equiv \int_0^1 d\mu \,
{d \mu_0 \over d \mu} \, \Bigl\{ 1 - \alpha r^{-1/2} - {1 \over 2}
\zeta (1 - \mu_0^2) \Bigr\}^{-1/2} \, . \eqno(2.24)$$
Notice that the density distribution in the spherical limit
is given by $\rho = C r^{-3/2}$ (see Shu 1977) so that the
effects of rotation and radiation pressure are incorporated
into the asphericity function ${\cal A} (r)$.

We can evaluate the integral appearing in equation [2.24] by changing
the integration variable from $\mu$ to $\mu_0$ and changing the lower
end of the range of integration from 0 to a critical value $\muck$.
The difference in the range of integration arises because streamlines
from all initial angles cannot fall to arbitrarily small radii.  For
large radii, streamlines from all initial angles (i.e., all values of
$\mu_0$) are represented.  Inside the centrifugal barrier $R_C$,
however, only streamlines originating preferentially from the poles
can reach these smaller radii (although for $\alpha \ne 0$, no
streamlines will reach radii $r < \alpha^2$).
The last streamline that can reach a given radius for a given $\alpha$
is given by $\muck$; clearly $\muck \to 0$ as $r \to \infty$ and
$\muck \to 1$ as $r \to 0$. After evaluating the integral, we find
$${\cal A} (r) =  \bigl( {2 \over \zeta} \bigr)^{1/2}
\log \Biggl[ { 1 + (2/\zeta)^{1/2} (1 - \alpha r^{-1/2})^{1/2}
\over
\muck + \Bigl[ (2 / \zeta) (1 - \alpha r^{-1/2}) + \muck^2 - 1
\Bigr]^{1/2} }  \Biggr] \, . \eqno(2.25)$$
The asphericity function ${\cal A}(r)$ is shown in Figure 4
for different amounts of radiation pressure. As the radiation
pressure is increased, the cusp near the centrifugal radius
becomes sharper and is also pushed back to larger radii.
Notice also that for $\alpha > 0$, a second ``cusp'' forms
at the location of the shock surface (as shown in Figure 2).

We can explicitly evaluate the asphericity function in the limits of
large and small radii.  For large radii, all of the streamlines
reach $r$ and hence $\muck = 0$. For this case, we have
$${\cal A} (r) =  \bigl( {2 \over \zeta} \bigr)^{1/2}
\log \Biggl[ { 1 + (2/\zeta)^{1/2} (1 - \alpha r^{-1/2})^{1/2}
\over \Bigl[ (2 / \zeta) (1 - \alpha r^{-1/2}) - 1
\Bigr]^{1/2} }  \Biggr] \, . \eqno(2.26)$$
Thus, in the limit $r \to \infty$, we have ${\cal A} (r) \to 1$.  In
other words, as expected, the departure of the solution from the pure
spherical case disappears in the limit of large radii.
In the opposite limit of small radii, the result depends on
the relative size of the centrifugal radius $R_C$ and the
radial scale $\alpha^2$ set by radiation pressure.  Under
the ordering
$$\alpha^2 \ll r \ll R_C \, , \eqno(2.27)$$
we obtain the result
$${\cal A} (r) \approx (2 - \sqrt{2}) {r \over R_C} \, ,
\eqno(2.28)$$
which is identical to the case of no radiation pressure (see ALS).
Thus, the density distribution has the form $\rho \sim r^{-1/2}$
at the intermediate spatial scales given by equation [2.27].
Notice that we need not consider the other limit in which
$\alpha^2 \ge r$ because all orbits will turn around before
reaching the radius $r$.

The above discussion implies that there is an effective centrifugal
radius in this problem.  This radius, denoted here as $R_{CR}$,
is the smallest radius reached by streamlines from all initial
angles (all values of $\mu_0$).  This radius is easily calculated
from the orbit solution and is given by
$$R_{CR} = R_C \, \Bigl[ 1 - \alpha R_C^{-1/2} (1 - \sqrt{2}/2)
\Bigr]^{-2} \, . \eqno(2.29)$$
Notice that, as expected, the effective centrifugal radius is
larger than for the case with no radiation pressure ($\alpha$
= 0).

\bigskip
\centerline{\it 2.6 Matching onto Outer Solutions}
\medskip

In the limit of large radius, this inner solution for the collapse
flow matches smoothly onto the general class of spherical infall
solutions which have a well defined mass infall rate; the isothermal
collapse solution (Shu 1977) is a well known example.  In this limit,
the orbit equation [2.13] shows that $\mu \to \mu_0$ and the velocity
field becomes purely radial with $v_r \to - (2GM/r)^{1/2}$ (see
equations [2.17 -- 2.19]).  Thus, in the limit of large $r$, the
density field becomes simply $\rho = C r^{-3/2}$, with the constant
$C$ defined by equation [2.23].  In other words, the density
distribution approaches the form appropriate for spherical free-fall
with a given (not necessarily constant) mass infall rate.

In order for the large radius limit to apply, the radius $r$ must
be large compared to both the centrifugal radius $R_C$ and the
radial scale $R_R = \alpha^2$ set by radiation pressure. Thus,
this solution approaches the ``large radius limit'' slightly
slower than the case of rotation only.  However, since we
generally expect $\alpha^2 \ll R_C$, this effect is small.

\bigskip
\centerline{\bf 3. MAXIMUM MASS SCALES FOR FORMING STARS}
\nobreak
\medskip

In this section, we discuss the maximum mass of a star that can be
formed through a rotating infall flow which includes radiation
pressure.  In the absence of radiation pressure, most of the infalling
material falls directly onto a circumstellar disk.  Although disk
physics is not well understood, stability considerations suggest that
a substantial fraction of this mass will eventually accrete onto the
central star.  When radiation pressure is important, some particle
orbits will be ``turned away'' before they impact the disk or the
star.  As radiation pressure increases, less and less of the infalling
material actually hits the disk.  When the radiation pressure reaches
a critical level, {\it none} of the infalling material can reach the
disk and the star/disk system effectively becomes isolated from its
environment.  In this section, we calculate the mass scale at which
this isolation occurs for several cases of interest.   If disk
accretion is efficient, then this mass scale is the maximum stellar
mass that can be formed within this infall scenario.

\bigskip
\centerline{\it 3.1 Mass Limits for the Infall Collapse Solution}
\medskip

Given the full solution for the particle orbits in the previous
section, we can determine conditions for which material hits the disk
and the corresponding conditions for which material becomes turned
away by radiation pressure.  We first note that parcels of gas will
hit the disk when they pass through the equatorial plane, i.e., when
$\mu$ = 0.  Using the orbit equation, we can solve for the radius at
which the particles hit the disk; this condition can be written
in the form
$$R_C (1-\mu_0^2 ) r^{-1/2} = R_C^{1/2} (1-\mu_0^2 )^{1/2}
+ \alpha (\sqrt{2}/2 - 1) \, . \eqno(3.1)$$
Similarly, using the solution for the velocity field, we can
determine that radius at which orbits will turn around due to
the radiation pressure; this condition can be written in the form
$$R_C (1-\mu_0^2 ) r^{-1/2} = - \alpha +
\Bigl[ \alpha^2 + 2 R_C (1-\mu_0^2 ) \Bigr]^{1/2} \, .
\eqno(3.2)$$
Note that, as formulated, the orbits will always turn around
for some sufficiently small radius.  This behavior is simply
a reflection of the fact that the repulsive term in the potential
(due to radiation pressure) always becomes larger than the
gravitational attraction for sufficiently small radius.
In practice, however, when the ``turn around radius'' becomes
smaller than the radius of the dust destruction front,
parcels of gas will not actually turn around.

Given the expressions [3.1] and [3.2] for the radii at which
infalling parcels of gas hit the disk and turn around, respectively,
we can write down a condition which must be met in order for
parcels of gas to hit the disk before they reach the turn around
radius. This condition can be written in the form
$$\alpha^2 < 2 R_C (1 - \mu_0^2) \, . \eqno(3.3)$$
Alternately, we can use this condition to determine the
range of initial angles which lead to orbits which
intersect the disk.  Formally, all particle orbits will
hit the disk provided that
$$\mu_0^2 < 1 - \alpha^2 / 2 R_C < 1 \, . \eqno(3.4)$$
Thus, when
$$\alpha^2 > 2 R_C \, , \eqno(3.5)$$
then {\it none} of the orbits will hit the disk plane and all of the
orbits will turn around.  In other words, when $\alpha$ becomes
greater than $2 R_C$, the infall flow onto the central star/disk
system effectively stops.  As a result, the condition [3.5]
defines the maximum mass of a star that can form through this
infall scenario.  This condition makes intuitive sense: The
quantity $\alpha^2$ defines the radius at which the radiation
pressure dominates over gravity; the centrifugal radius $R_C$
defines the radius at which angular momentum dominates over
gravity.  In order for radiation pressure to dominate over
angular momentum, some condition like equation [3.5] is necessary.

In order to evaluate the maximum mass scale, we must evaluate the
coupling parameter $\alpha$ (see equation [2.7]).  Thus, we must
find the dust destruction radius as a function of the stellar
luminosity.  We find that
$$r_d = \Biggl[ { L \over 16 \pi \sigma T_d^4 }
{\kappa_\ast \over \kappa_P (T_d) } \Bigr]^{1/2} \,
\equiv R_D \, {\widetilde L}^{1/2} \,
\approx 5.7 \times 10^{11} {\rm cm} \, \, \,
{\widetilde L}^{1/2} \, , \eqno(3.6)$$
where ${\widetilde L}$ is the total luminosity in units of $L_\odot$
and where $\kappa_\ast$ is the opacity of the material to stellar
photons (see Stahler, Shu, \& Taam 1980 and Appendix B of Adams 1990).
We have defined a fiducial value $R_D$ of the dust destruction radius
which is that appropriate for the luminosity of the sun.
Notice that the dust grains absorb stellar photons with an efficiency
proportional to $\kappa_\ast$ and emit photons with a lower efficiency
proportional to $\kappa_P(T_d)$.  Thus, the dust grain temperature at
any given radius is larger than the naive ``planet approximation'' by
the factor $[\kappa_\ast/\kappa_P (T_d)]^{1/4}$ $\sim$ 1.6, where we
have adopted $\kappa_\ast$ = 200 cm$^2$/g.

Using the expression [3.6] for the dust destruction radius, we can
evaluate the coupling parameter $\alpha$ and then find the maximum
mass scale.  We first consider the case of isothermal initial
conditions, since most previous work on star formation has begun with
the collapse of an isothermal cloud core; however, the alternate
``logatropic'' initial state described below might be more
applicable to the formation of high mass stars. For the isothermal
case, after considerable rearrangement, we can write the condition [3.5]
for radiation pressure to halt infall onto the disk in the form
$$L^{5/2} \, M^{-5} \ge 2 [3 \pi c \Omega]^2 \, G^5 \,
a^{-8} \, [\kappa_P(T_d)]^{-3/2} \, \kappa_\ast^{-1/2} \,
[\pi \sigma T_d^4]^{1/2} \, . \eqno(3.7{\rm a})$$
Inserting numerical values and rearranging once again, we find
$${\widetilde L} \, \, {\widetilde M}^{-2} \, \ge 1570
\, \, {\widetilde \Omega}^{4/5} \, \, a_{35}^{-16/5}
\, , \eqno(3.7{\rm b})$$
where we have defined ${\widetilde \Omega} \equiv \Omega$/(1 km s$^{-1}$
pc$^{-1}$) and $a_{35} \equiv a$/(0.35 km s$^{-1}$).  The relation [3.7]
implies a maximum mass of a forming star for a given set of the initial
conditions.

Notice that the mass $M$ appearing in these constraints is the
{\it total} mass that has fallen onto the star/disk system. In
general, only a fraction of this mass becomes part of the star
so that we should write $M_\ast = \gamma M$.  However, disk
stability considerations (Adams, Ruden, \& Shu 1989; Shu et al.
1990) show that $\gamma \ge 2/3$; otherwise, the disk would be
violently unstable to self-gravitating perturbations.  Thus, the
difference between $M$ and $M_\ast$ is not overly large.

In order to evaluate this limit, we must use some type of
mass/luminosity relation.  The stars of interest are generally in the
mass range of 20 -- 100 $M_\odot$ so that their evolutionary time
scales are short compared to the infall time scale.  This ordering
implies that the stars will have (nearly) their main sequence
configurations while they are still in the infall phase of evolution
(see, e.g., Yorke \& Kr{\"u}gel 1977; Yorke 1979; Palla \& Stahler
1990).  We can thus use a main sequence mass/luminosity relation for
this present discussion (e.g., Ezer \& Cameron 1967; Allen 1976;
Phillips 1994).

Our results for an isothermal collapse are shown in Figure 5 where the
mass limit is given as a function of the initial conditions $a$ and
$\Omega$.  For a given mass, the region of parameter space to the
lower right of the curve is excluded, i.e., a star of the given mass
cannot form with initial conditions in that region.  The most
restrictive limit occurs for stars with the largest value of $L/M^2$
(see equation [3.7b]) which occurs for $M \approx 50 M_\odot$.  Thus,
if the rotation rate is large enough and/or the effective sound speed
is low enough, then the luminosity never becomes large enough to halt
the infall (in other words, the left hand side of equation [3.7b]
never exceeds the right hand side). In this case, radiation pressure
alone does not restrict the mass of a forming star.

Next, we consider the case of logatropic initial conditions.  In this
case, the condition for radiation pressure to halt the infall can be
written in the form
$$L^{5/2} \, M^{-3} \ge {16 \over \pi P_0} \, [3 \pi c G \Omega]^2 \,
[\kappa_P(T_d)]^{-3/2} \, \kappa_\ast^{-1/2} \, [\pi \sigma T_d^4]^{1/2}
\, . \eqno(3.8{\rm a})$$
After inserting numbers and rearranging, we obtain
$${\widetilde L} \, {\widetilde M}^{-6/5} \, \ge \,
1.1 \times 10^4 \, {\widetilde \Omega}^{4/5} \,
{\widetilde P}^{-2/5} \, ,
\eqno(3.8{\rm b})$$
where ${\widetilde \Omega}$ is defined as before and where we
have also defined ${\widetilde P} \equiv P_0$ /
($3.6 \times 10^{-11}$ dyne cm$^{-2}$); this numerical value
for the pressure scale corresponds to the magnetic pressure
$B^2/8 \pi$ for a field strength of 30$\mu$G.  In order to
evaluate this limit, we can once again use the main sequence
mass/luminosity relationship.

The mass limit as a function of logatropic initial conditions is shown
in Figure 6.  As before, for a given mass, the region of parameter
space to the lower right of the curve is excluded, i.e., a star of the
given mass cannot form with initial conditions in that region.  Also,
the quantity ${\widetilde L} {\widetilde M}^{-6/5}$ has a maximum
value close to $10^4$; thus, if the pressure scale $P_0$ is
sufficiently small and/or the rotation rate is sufficiently large,
then radiation pressure alone cannot restrict the mass of the forming
star.  Notice also that the mass limits for these logatropic initial
conditions are generally less restrictive than the isothermal case
shown in the previous figure.  In other words, for the expected range
of pressure scales, a lower rotation rate is required to allow
material to fall to sufficiently large radii in the disk to evade the
effects of radiation pressure.

As shown above, the maximum mass of a forming star increases with
increasing rotation rate $\Omega$ and decreases with increasing
pressure of the initial state (given by either the effective sound
speed $a$ or the pressure scale $P_0$).  This behavior has a
straightforward physical interpretation: When the rotation rate is
larger, the angular momentum is larger, and infalling material tends
to fall to larger radii in order to conserve angular momentum.  As a
result, material does not fall as far down the potential well and
cannot be affected (as much) by radiation pressure.  When the sound
speed (or the pressure scale) for the initial configuration is larger,
the infall rate $\dot M$ is larger, and material falls farther down
the potential well for a given rotation rate; at these smaller radii,
radiation pressure has a greater effect.  The difference between the
isothermal and logatropic initial conditions is also clear: The latter
case has more material at larger radii in the initial equilibrium
state and hence the effects of angular momentum conservation are
greater for a given rotation rate; this effect, in turn, makes the
mass limits weaker for the logatropic case.

\bigskip
\centerline{\it 3.2 Mass Limit for Spherical Infall}
\nobreak
\medskip

For comparison, we write down the maximum mass scale for the case
of spherical infall (no rotation). In this case, the parcels
of gas and dust fall inward radially through the potential well
given by equation [2.6].  The orbits will ``turn around'' if the
coupling term $\alpha r^{-1/2}$ becomes larger than unity
before the parcel reaches the dust destruction radius (where
the coupling vanishes). Thus, the condition for radiation
pressure to dominate is given by the relation
$$\alpha^2 \ge r_d \, , \eqno(3.9)$$
where the dust destruction radius is given by equation [3.6] above.
Notice that the condition for reversing spherical infall is
generally much weaker than the condition for reversing the
rotating infall (because, in general, $R_C \gg r_d$).  Thus,
the maximum mass scale for the spherical case will be much
lower than the more realistic rotating case found above.
After some rearrangement, we can write this condition in the form
$${L \over M} \ge {6 \pi G c \over \kappa_P(T_d)}
\, . \eqno(3.10{\rm a})$$
Using numerical values, we find
$${ {\widetilde L} \over {\widetilde M}} \ge 640
\, . \eqno(3.10{\rm b})$$
This condition has the following simple physical interpretation.
For sufficiently large ${\widetilde L}/{\widetilde M}$, the specific
work done on the infalling gas by the radiation field exceeds the
depth of the gravitational potential well at $r_d$ (see ALS).
This value is essentially equivalent to that found by Kahn (1974),
although he used a higher value for the dust destruction
temperature (3670 K) and found a corresponding higher value
for the maximum mass (see also Wolfire \& Cassinelli 1987 for
further discussion).

\bigskip
\centerline{\it 3.3 General Mass Limits}
\medskip

We have shown that the condition for radiation pressure
to reverse infall onto a star/disk system is that the
turnaround radius $R_R$ of the radiation pressure must be
larger than the centrifugal radius $R_C$ of the rotating infall.
In order words, the condition
$$R_R > \eta \, R_C \eqno(3.11)$$
defines the maximum mass of a forming star; the dimensionless
constant $\eta$ is of order unity, but its exact value must be
determined from the true infall solution.
In this subsection, we use this result to study the effects
of varying the different parameters in the problem.  In
particular, we find the maximum mass of a forming star
for different types of coupling between the radiation field
and the infalling gas and for different initial conditions
in the cloud core.

We first define a radiation pressure ``structure constant''
$\delta_R$ through the relation
$$\delta_R \equiv { L_\odot \kappa_P [T_d] \over
4 \pi G c M_\odot } \approx {1 \over 430} \, . \eqno(3.12)$$
Thus, $\delta_R$ is an intrinsically small parameter in the
problem.

We want to consider different power-law forms for the radiation
pressure force (see equation [2.5]).  We can write this force
in the form
$$f = {G M \over r^2} \,  \delta_R \,
({\widetilde L} / {\widetilde M}) \Bigl( {r_d \over r}
\Bigr)^q \, , \eqno(3.13)$$
where $q$ is an arbitrary power-law index which we expect to lie
in the range $1/3 \le q \le 1$.  The effective turnaround radius
$R_R$, where radiation pressure dominates gravity, is given by
$$R_R = \Bigl[  \delta_R \, \, {\widetilde L} / {\widetilde M}
\Bigr]^{1/q} \, r_d \, . \eqno(3.14)$$
In order to isolate the effects of different power-law
indices $q$, we specify the initial state to be in logatropic
equilibrium so that the centrifugal radius is given by
equation [2.15].  We can write this radius in the form
$$R_C = R_0 \, \, {\widetilde M} \, {\widetilde \Omega}^2 \,
{\widetilde P}^{-1} \, \approx 9 \times 10^{15} {\rm cm} \,
{\widetilde M} \, {\widetilde \Omega}^2 \, {\widetilde P}^{-1}
\, , \eqno(3.15)$$
where $\widetilde M$, ${\widetilde \Omega}$, and ${\widetilde P}$ are
defined as before.  The radius $R_0$ is the ``natural value'' of
the centrifugal radius.

For this case, the mass limit can be written in the form
$$[\delta_R {\widetilde L} / {\widetilde M}]^{1/q} \,
{\widetilde L}^{1/2} \, {\widetilde M}^{-1} \, > \,
\eta \, {R_0 \over R_D} \, {\widetilde \Omega}^2 \, {\widetilde P}^{-1}
\approx 1.5 \times 10^4 \, \eta \, {\widetilde \Omega}^2 \,
{\widetilde P}^{-1} \, . \eqno(3.16)$$
For a given set of initial conditions, the right hand side of
this equation is a (generally large) constant.  The left hand side
is an increasing function of the mass, although it saturates at
sufficiently large masses.  As the index $q$ increases, the
mass scale given by equation [3.16] also increases. In other words,
a larger mass star is needed to reverse the infall. This result is
expected physically: As the index $q$ increases, the effective range
of the radiation pressure force decreases (see equation [3.13])
and hence radiation has a smaller effect on the infall.

We also want to determine the effects of different initial
conditions on the maximum mass scale for forming stars.
As discussed in \S 2, we expect the initial equilibrium states
for star formation to be described by equations of state which
are isothermal or {\it softer}.  As a result, the initial
mass profile should be of the form $M \sim r^b$ where the
index $b$ lies in the range $1 \le b \le 2$.  This set of
states implies a corresponding set of centrifugal radii of
the form
$$R_C = R_0 {\widetilde M}^d \, \eqno(3.17)$$
where the index $d$ lies in the range $1 \le d \le 3$ (see equations
[2.14] and [2.15]).  The fiducial radius $R_0$ must be close to the
disk radius for solar type stars, i.e., $R_0 \sim$ 100 AU. The mass
limit for forming stars can then be written in the form
$$[\delta_R {\widetilde L} / {\widetilde M}]^{1/q} \,
{\widetilde L}^{1/2} \, {\widetilde M}^{-d} \, > \,
\eta \, {R_0 \over R_D} \,
\, . \eqno(3.18)$$
Typically, we expect the right hand side of this constraint
to be a large number, $\sim 10^3 - 10^5$.  As the index $d$
increases, the constraint becomes harder to satisfy, i.e.,
a larger mass star is required to reverse the infall.

\bigskip
\centerline{\bf 4. ADDITIONAL ISSUES}
\medskip

In this section, we briefly and heuristically discuss several
additional issues which are relevant to infall collapse solutions
of the type considered in this paper.  In particular, we discuss the
leading order effects of magnetic fields, a critical value of the mass
infall rate, and the manner in which rotation places additional
constraints on the infall solution and the corresponding maximum
mass scales of forming stars.

\bigskip
\centerline{\it 4.1 Magnetic Fields}
\medskip

Thus far in this discussion, we have ignored the effects of magnetic
fields on the collapse.  However, since molecular clouds are supported
on large scales by magnetic fields (Zuckerman \& Palmer 1974; Shu et
al. 1987) and since cloud cores are most likely formed through the
process of magnetic field diffusion (e.g., Mouschovias 1978; Shu 1983;
Nakano 1984; Lizano \& Shu 1989), magnetic fields should be
incorporated into the collapse solution simultaneously with rotation
and radiation pressure.  We can get an order of magnitude estimate of
the magnetic field strength required to affect the infall as follows.
The effective strength of the field can be measured by the radius
$R_B$ at which the perturbation in the meridional velocity produced by
the Lorentz force is equal to the asymptotic free-fall velocity; this
radius plays a role in the collapse flow similar to that of the
centrifugal radius and thus defines the outer radius of a ``Pseudodisk''
(see Galli \& Shu 1993ab).   The magnetic radius can be written
$$R_B = k_B \Bigl[ {G^2 B^4 \over a } \Bigr]^{1/3} t^{7/3}
\, , \eqno(4.1)$$
where $k_B$ is a dimensionless constant.
For the case of an isothermal collapse model, we can replace the
time variable by the mass, i.e., $t = M/{\dot M}$ = $MG/a^3 m_0$.
The magnetic radius is then given by
$$R_B = k_B \Bigl[ {G^9 B^4 M^7 \over a^{22} \, m_0^7 } \Bigr]^{1/3}
\, , \eqno(4.2)$$
where $m_0$ = 0.975 is the dimensionless mass
determined from the isothermal collapse solution (Shu 1977).

Since the magnetic radius plays an analogous role to the
centrifugal barrier which we have already included in the
problem, we can make the following rough argument: We expect
radiation pressure to reverse the infall for a magnetically
collapsing core when the turnaround radius $R_R$ exceeds the
magnetic radius $R_B$.  This condition can be written in the
form
$$L^{5/2} \, M^{-13/3} \ge \, \eta k_B \, [12 \pi c]^2 G^5
\, \, \Bigl[ {B^4 \over a^{22} \, m_0^7 } \Bigr]^{1/3} \, \,
[\kappa_P(T_d)]^{-3/2} \, \kappa_\ast^{-1/2} \,
[\pi \sigma T_d^4]^{1/2} \, ,
\eqno(4.3{\rm a})$$
where we have included a constant $\eta$ of order unity.
After inserting numerical values, we can write this result
in the form
$${\widetilde L} \, {\widetilde M}^{-26/15} \ge 1.9 \times 10^4 \,
{\widetilde B}^{8/15} \, a_{35}^{-44/15} \, , \eqno(4.3{\rm b})$$
where we have defined ${\widetilde B} \equiv B$/30$\mu$G and
where we have set $\eta = 1 = k_B$ to obtain the numerical value.
This relation defines the maximum mass of a forming star for a
collapse flow which includes magnetic fields and radiation pressure.
This maximum mass scale is a function of the initial conditions,
defined here by the sound speed $a$ and the magnetic field
strength $B$; the result is shown in Figure 7.

\bigskip
\centerline{\it 4.2 Critical Mass Infall Rate}
\medskip

The approximations of this paper become invalid when
the mass infall rate becomes too small.
In order for infall to occur, the mass infall rate
must be large enough to produce a ram pressure greater
than the radiation pressure at the dust destruction front.
This condition can be written in the form
$${\dot M} v_{in} > L/c \, , \eqno(4.4)$$
where $v_{in}$ is the speed of the infalling material
evaluated at the radius $r_d$. This condition implies
that for sufficiently high luminosity $L$, the radiation
pressure will dominate the total ram pressure and will
terminate the infall.  However, this constraint is not
overly restrictive.  If we insert numerical values, the
critical mass infall rate, denoted here as ${\dot M_C}$,
becomes
$${\dot M_C} = 1.4 \times 10^{-10} M_\odot \, {\rm yr}^{-1} \,
{\widetilde L}^{5/4} \, {\widetilde M}^{-1/2} \, , \eqno(4.5)$$
where we have used equation [3.6] to evaluate the dust destruction
radius $r_d$.
In order for infall to occur, the mass infall rate must exceed
this critical rate.  For example, for a massive star with
${\widetilde M} = 100$, the luminosity is ${\widetilde L}$
= $1.2 \times 10^6$ and the critical mass infall rate is
${\dot M_C}$ = $6 \times 10^{-4}$ $M_\odot$ yr$^{-1}$.

\bigskip
\centerline{\it 4.3 Rotation and Time Constraints}
\medskip

In obtaining the maximum masses of forming stars (\S 3), we have
assumed that infalling material which reaches the disk can eventually
become part of the star through the process of disk accretion.
Although disk physics is not completely understood, stability
considerations suggest that this assumption is reasonable,
provided that sufficient time is available for disk accretion
to occur. In this section we show that the rather short lifetime
of massive stars places an additional constraint on this process.

Conservation of angular momentum causes incoming material to fall
initially onto a circumstellar disk rather than directly onto the
star.  In the absence of disk accretion, this effect shuts off the
mass flow onto the star.  However, even in the presence of disk
accretion, angular momentum can eventually shut off the infall because
the time scale for disk accretion to take place becomes longer than
the lifetime of the star.  Very roughly, the shortest time scale for a
disk accretion process is the dynamical time scale at the outer disk
edge.  For a disk with radius $R_C$, the centrifugal radius,
this time scale is roughly given by
$$\tau^2 = {R_C^3 \over G M} \, , \eqno(4.6)$$
where we have assumed that the outer disk rotation curve is nearly
Keplerian. If we now enforce the constraint that this time scale $\tau$
must be shorter than the lifetime $\tau_\ast$ of a massive star,
we obtain a constraint on the centrifugal radius of the form
$$R_C \le \tau_\ast^{2/3} \, (G M)^{1/3} \, . \eqno(4.7)$$
We note that this constraint is necessary but not sufficient.
A very massive star can significantly disrupt a circumstellar
disk and thereby disrupt disk accretion and evolution (see
Hollenbach et al. 1994).

We can use the above result to obtain a maximum stellar
mass constraint that is independent of the initial conditions.
We have already shown that in order for radiation pressure
{\it not} to reverse the infall, we must have $R_R = \alpha^2$
$< 2 R_C$.  Thus, a star can continue to gain mass through the
combination of infall and disk accretion only when following
ordering is satisfied,
$$\alpha^2 / 2 \, < \, R_C \, < \,
\tau_\ast^{2/3} \, (G M)^{1/3} \, . \eqno(4.8)$$
Thus, when the radiation turnaround radius ($R_R \sim \alpha^2$)
exceeds the critical radius defined by the right hand side of this
inequality, the ordering constraint cannot be met for {\it any}
set of initial conditions (for any collapse model, isothermal
or otherwise).  Thus, the condition for the maximum stellar mass
can be written in the form
$$L^{5/2} \, M^{-7/3} = 2 \tau_\ast^{2/3} \,
G^{7/3} [12 \pi c]^2 \,
[\kappa_P(T_d)]^{-3/2} \, \kappa_\ast^{-1/2} \,
[\pi \sigma T_d^4]^{1/2} \, .
\eqno(4.9{\rm a})$$
As usual, we can insert numerical values and rearrange
this limit to obtain the form
$${\widetilde L} \, {\widetilde M}^{-14/15} = 5.6 \times 10^4 \,
\tau_6^{4/15} \, , \eqno(4.9{\rm b})$$
where we have defined $\tau_6 \equiv \tau_\ast/(10^6$yr).
As a reference point, note that $\tau_6 \approx 3.6$ for
a star with mass ${\widetilde M}$ = 100 (Ezer \& Cameron 1967).
The limit [4.9] represents the largest mass accessible to a star that
forms within the collapse scenario considered here for any set
of initial conditions.  The left hand side of this equation
is (almost) the luminosity to mass ratio in solar units;
for massive main sequence in the mass range 100 -- 200 $M_\odot$,
this ratio is approximately 1--2 $\times 10^4$, i.e., slightly
smaller than the numerical value on the right hand side.
Thus, the maximum luminosity to mass ratio allowed for
forming stars within this scenario is slightly larger
than the $L/M$ ratios for very massive stars.

For completeness, we note that the time constraint considered
here can be written as a mass limit which is independent of
the effects of radiation pressure. For isothermal initial
conditions, this limit can be written in the form
$$M < \, {\sqrt 8} \,
G^{-1} \, \tau_\ast^{1/4} \, a^3 \, \Omega^{-3/4} \, ,
\eqno(4.10{\rm a})$$
or, in terms of dimensionless quantities,
$${\widetilde M} < 30 \, {\widetilde \Omega}^{-3/4} \,
a_{35}^3 \, \tau_6^{1/4} \, . \eqno(4.10{\rm b})$$
For a given set of initial conditions $a$ and $\Omega$,
the mass of a star that can form through the combined
process of infall and disk accretion is limited by the
constraint of equation [4.10].  Notice that this limit
only becomes restrictive for high mass stars because of
the dependence on the stellar lifetime $\tau_6$.

Finally, we also note that there exists yet another a related
constraint on the mass.  In the original molecular cloud core,
gaseous material at sufficiently large radii is centrifugally
supported in the initial state.  This radius is given by
$R = a/\Omega$ for an isothermal cloud core. Thus, the total
mass that is available to fall inward is given by
$$M = {\cal F}  {2 a^3 \over G \Omega} \,  \eqno(4.11{\rm a})$$
where we have used the mass profile for an isothermal
cloud core and where we have introduced a factor $\cal F$
to take geometry into account.  Gas located in the polar
directions is not centrifugally supported and can thus fall
inwards.  We thus obtain
$${\widetilde M} = 21 {\cal F} \, a_{35}^3 \,
{\widetilde \Omega}^{-1} \, . \eqno(4.11{\rm b})$$
The reason why these two constraints [4.10] and [4.11] are
nearly identical is that the time scale for stellar evolution of
massive stars (about $10^6$ yr) is about the same as the rotation
period of a molecular cloud core.

\newpage
\bigskip
\centerline{\bf 5. SUMMARY AND DISCUSSION}
\nobreak
\medskip

In this paper, we have generalized the infall collapse solution
for star formation to include the effects of radiation pressure
in the inner regime.  Our results can be summarized as follows:

{\bf [1]} We have shown the effects of radiation pressure can be
modeled using a modified potential.  The radiation field is expected
to be nearly spherically symmetric and creates an outward force.
Furthermore, the radial dependence of the radiation pressure force
is expected to decrease with radius faster than the gravitational
force.  As a result, the effects of radiation pressure can be
incorporated using a modified potential of the general form
$$V_{\rm eff} = - {G M \over r} \Bigl\{ 1 - g(r) \Bigr\} \, , $$
where $g(r)$ is a decreasing function of radius $r$.  We have used
the approximation $g = \alpha r^{-1/2}$ which corresponds to
assuming that the coupling between the radiation field and
the gas is proportional to the Planck mean opacity and that
the temperature distribution has the simple form $T \sim r^{-1/2}$.

{\bf [2]} We have found an analytic solution (equation [2.13]) for
the zero energy orbits of the modified potential described above (see
equation [2.6]).  This solution provides the velocity field (equations
[2.17 -- 2.19]) and the density field (equation [2.21]) for infall
collapse solutions which simultaneously include both angular momentum
and radiation pressure.

{\bf [3]} The infall solution has three important radial scales: The
dust destruction radius $r_d$, the centrifugal radius $R_C$, and the
radius $R_R = \alpha^2$ at which radial orbits are reversed by
radiation pressure. The characteristics of the orbits and the infall
solution are determined by the ratios of these three radial scales.
In order for angular momentum to play a significant role in the
collapse, the centrifugal radius $R_C$ must exceed the dust
destruction radius $r_d$.  Similarly, in order for radiation pressure
to play an important role, the turnaround radius $R_R$ must exceed
the dust destruction radius $r_d$.  For the case $R_C \gg R_R$,
radiation pressure perturbs the orbits so that infalling material
hits the disk at larger radii; however, the qualitative nature of
the infall is similar to the case of no radiation pressure.  In the
opposite limit $R_R \gg R_C$, the orbits turn around before they
impact the disk and infall effectively ceases.

{\bf [4]} This infall collapse solution implies a maximum mass scale
for forming stars.  For this solution, the condition for radiation
pressure to halt the infall onto the central star/disk system is given
by $\alpha^2 = R_R > 2 R_C$.  The resulting maximum mass scale is a
function of the initial conditions for protostellar collapse (see
Figures 5 and 6).  In general, the constraints on the masses of
forming stars (see equations [3.7] and [3.8]) are considerably less
restrictive for rotating collapse with radiation pressure than for the
case of purely spherical infall (equation [3.10]).  Our results show
that massive stars with $M_\ast \sim$ 100 $M_\odot$ can form for a
wide range of initial conditions.

{\bf [5]} We have argued that the constraint which limits infall
onto the disk, $R_R > R_C$, is more robust than its derivation.
As a result, we can use this condition to find the mass limits
on forming stars for a wide range of initial conditions and
for different assumptions about the coupling of the radiation
field to the infalling gas (see \S 3.3).  As the coupling of
the radiation field falls off more quickly with radius, the
effects of radiation pressure decrease, and stars of higher
mass can form with a given set of initial conditions. Similarly,
as the equation of state becomes softer, stars of higher mass
can form.  These statements are quantified by equations [3.16]
and [3.18].

{\bf [6]} We have explicitly found orbit solutions for different
assumptions about the coupling between the radiation field and the
infalling gas.  In particular, we have considered the two limiting
cases in which the strength of the coupling decreases as slowly
and as rapidly as possible (see Appendix B).

{\bf [7]} We have discussed the leading order effects of magnetic
fields on a rotating collapse which includes radiation pressure.
The net effect of magnetic fields is to prevent material from
falling as far inwards (due to the Lorentz force). As a result,
radiation pressure has {\it less} of an effect in the presence
of magnetic fields.

{\bf [8]} We have shown that this scenario for forming massive
stars (through the combination of infall and disk accretion) breaks
down when the natural time scale of the disk becomes longer than
the lifetime of a massive star.  This effect places a limit on the
ratio of luminosity to mass ($L/M$) for forming stars.  Although
this limit is independent of initial conditions, it is not very
restrictive: Stars with masses in the range 100 -- 200 $M_\odot$ are
allowed by a small margin (see equation [4.9] and \S 4.3).

In some sense, the results of this paper show that radiation pressure
has less impact on the infall than previous studies (with spherical
infall) have implied. The basic physical reason for this difference is
that previous studies have not simultaneously included both angular
momentum and radiation pressure.  For the typical range of parameters
applicable to protostellar collapse, angular momentum plays an
extremely important role.  In particular, the centrifugal radius $R_C$
(which determines the size of the angular momentum barrier) is much
larger than the radius of the star or the dust destruction front.  As
a result, in order for radiation pressure to have a significant
impact, it must affect infalling material at the (large) size scales
of the centrifugal radius $R_C$.

The results of this paper can be tested observationally, although
such tests are difficult. The collapse solution of
this paper predicts that the infalling envelopes of forming massive
stars will be highly evacuated for radii less than the turnaround
radius $R_R$, and will have the general form as illustrated in
Figures 2 and 3. Although the density distribution of circumstellar
material is not directly measurable, its form can be deduced (or at
least highly constrained) by comparing radiative transfer calculations
with observed spectral energy distributions and emission maps.
For high mass stars, some work along these lines has already
been done (Churchwell, Wolfire, \& Wood 1990; see also Wolfire \&
Churchwell 1994).  The former study shows
that the dust envelope around a newly formed O6 star is highly
evacuated within $\sim 10^{17}$ cm of the star and that the density
distribution must be fairly flat, i.e., the density must increase
inwards less rapidly than $\rho \sim r^{-1/2}$.  These results are
in basic agreement with those of this paper. For this star (with
mass $M=$ 34 $M_\odot$ and $L = 2.5 \times 10^5 L_\odot$), the
turnaround radius $R_R = 4 \times 10^{16}$ cm and the maximum
centrifugal radius consistent with the stellar lifetime is
$\sim 2 \times 10^{18}$ $\tau_6^{2/3}$ cm.  Thus, a circumstellar
envelope with a fairly flat density distribution ($\rho \sim r^{-1/2}$)
in the range $r = 10^{17} - 10^{18}$ cm arises naturally from our
collapse solution for massive stars.

Another testable prediction of this calculation is the presence of
rather large amounts of mass (10 -- 100 $M_\odot$) entering the
circumstellar disks associated with high mass stars.  For this
collapse solution, most of the infalling material falls directly onto
the disk rather than directly onto the star.  If the disk accretion
rate is at all slower than the envelope infall rate, then material
piles up to form a massive disk.  Such massive disks will have bright
dust continuum emission at millimeter and submillimeter wavelengths.
In addition, the ionized (inner) edges of the disks will be bright
centimeter wave sources.  Existing studies have searched for massive
disks associated with embedded protostars of low and intermediate
mass; some studies provide supporting evidence for massive disks
(e.g., Reipurth et al. 1993; Ho, Terebey, \& Turner 1994),
whereas other studies suggest that massive disks are somewhat rare
(e.g., Terebey, Chandler, \& Andr{\' e} 1993;
Andr{\' e} \& Montmerle 1994).

Although this paper provides a significant generalization of the
protostellar collapse problem, many unresolved issues still remain.
This work has been limited to gravitational potentials which
correspond to point masses, whereas realistic systems include
both extended components (i.e., disks) and time varying
components (i.e, binaries). These issues should be addressed
in future work.

\vskip 0.5truein
\centerline{Acknowledgements}
\medskip

We would like to thank Frank Shu, Joel Tohline, and Rick Watkins
for useful discussions.  We also thank the referee -- Sue Terebey --
for many useful suggestions.
This work was supported by an NSF Young Investigator Award,
NASA Grant No. NAG 5-2869, and by funds from the Physics Department
at the University of Michigan; we also thank the Institute for
Theoretical Physics at U. C. Santa Barbara and NSF Grant No.
PHY94-07194.

\vskip 0.5truein
\newpage
\bigskip
\centerline{\bf APPENDIX A: ALTERNATE FORM OF THE ORBIT EQUATION}
\nobreak
\medskip

In this Appendix, we briefly describe another way to express
the orbit equation.  This form is useful for evaluating
the various expressions found in the text. We first define
$$x = {\mu \over \mu_0} \, , \eqno({\rm A}1)$$
and
$$A \equiv { 2 [ \zeta r^{1/2} (1 - \mu_0^2) + \alpha ]^2 \over
2 \zeta r (1 - \mu_0^2) + \alpha^2 } - 1 \, . \eqno({\rm A}2)$$
Notice that the quantity $A=1$ at the turnaround radius
$\alpha^2 = R_R$.
The orbit equation can then be written in the form
$$x = A \cos{\widetilde \phi} +
[ 1 - A^2 ]^{1/2} \sin{\widetilde \phi} \, , \eqno({\rm A}3)$$
where $\cos{\widetilde \phi}$ is defined by equation [2.12] in
the text.  The expression [A3] gives $\mu$ as a function of the
initial value $\mu_0$. To obtain this form, we solved a quadratic
equation which formally has two roots; the second root,
which corresponds to using a minus sign in equation [A3],
is not relevant.

\newpage
\bigskip
\centerline{\bf APPENDIX B: LIMITING CASES FOR}
\centerline{\bf THE RADIATION PRESSURE APPROXIMATION}
\medskip

In this Appendix, we assess the severity of the approximation
used in this paper to model the effects of radiation pressure.
We have assumed that the radiation pressure can be incorporated
using the form for the coupling given by equation [2.2].
In order to evaluate this term, we have made two approximations.
The first is that we can use the Planck mean opacity for the
weighted mean opacity $\kappa_E$ and that the Planck mean
opacity is {\it linear} in temperature. The second approximation
is that we assume the temperature distribution has the simple
power-law form $T \sim r^{-1/2}$ for the radii of interest in
this problem.  Thus, the coupling term $\sim r^{-1/2}$ in this
approximation.  In general, we can write the effective potential
in the form
$$V_{\rm eff} = - {G M \over r} \Bigl\{ 1 - g(r) \Bigr\} \,
\eqno({\rm B}1)$$
where $g(r)$ is an unspecified dimensionless function which describes
the radial variation of the radiation pressure term relative to
the gravitational potential.  We expect that the function $g(r)$
will vary faster than a constant ($g=$ {\sl constant} corresponds
to a constant ratio of radiation pressure to gravity) but will
vary less quickly than $g \sim r^{-1}$.  The case considered in
the text has the intermediate behavior $g \sim r^{-1/2}$.

As a start, we can solve for the orbits in the two limiting cases
which bracket the expected behavior for the radiation pressure term.
In order to solve the orbit equation, we must solve the integral,
$$J[g] = \Delta \phi = \gamma^{1/2} \int_r^{\infty} {dr \over r}
\Bigl[ 2 r (1 - g) - \gamma \Bigr]^{-1/2} \, \eqno({\rm B}2)$$
where $\gamma = R_C (1 - \mu_0^2)$ is proportional to the square of
the specific angular momentum of a parcel of gas beginning at angle
$\theta_0$. In the first case, where the coupling term varies slowly
as possible, we take $g$ = {\sl constant} and obtain the solution
$$\zeta (1 - \mu_0^2) = (1 - \mu /\mu_0) (1 - g) \, .
\eqno({\rm B}3)$$
This solution corresponds to reducing the gravitational
constant by the factor $(1-g)$.
In the opposite limit, where the coupling varies as quickly
as possible, we write the function $g$ in the form
$g = \xi/r$.  After some algebra, this solution can be
written in the form
$$1 - \cos \Bigl[ \phi (1 + 2 \xi / \gamma)^{1/2} \Bigr] =
{2 \xi + \gamma \over r} \, , \eqno({\rm B}4)$$
where $\cos\phi$ = $\mu / \mu_0$.  For any case, the
orbital behavior should be bracketed by the solutions
given by equations [B3] and [B4].

We now consider another way to look at the problem of different
radial dependences for the coupling term $g(r)$.  If we change
variables according to
$$x^2 \equiv {\gamma \over 2 r} \, , \eqno({\rm B}5)$$
the integral $J[g]$ can be written in the form
$$J[g] = 2 \int_0^{x_e} \, dx \, [1 - g - x^2]^{-1/2}
\, , \eqno({\rm B}6)$$
where we have taken the limits of integration to be spatial infinity
(where $r \to \infty$ and $x \to 0$) and $x_e$ = $\sqrt{2}/2$, which
corresponds to the radial position ($r = \gamma$) at which a parcel
of gas hits the disk in the case of no radiation pressure ($g=0$).
We want to determine the effects of relatively small departures from
the case of no radiation pressure. We thus take the first variation
of the functional $J[g]$ about the ``point'' $g=0$ to obtain
$$\delta J = \int_0^{x_e} dx \, g \, [1 - x^2 ]^{-3/2} \, .
\eqno({\rm B}7)$$
We now consider a class of models for $g$ of the form
$$g = \beta (x/x_e)^n \, , \eqno({\rm B}8)$$
where the index $n$ is confined to the range $0 - 2$
(which corresponds to the expected radial behavior of
$g(r)$ as described above).  We can evaluate the first
variation $\delta J$ for the various models of the coupling
given by equation [B8].  We find that $\delta J/\beta$ =
1, $\sqrt{2} - 1$, and $1 - \pi/4$ for the cases $n$ = 0, 1, and 2,
respectively.  Thus, the exact form of the coupling does not
greatly change the size of the action for a given value of
$\beta$.  In other words, the functional form of the coupling
$g$ only changes the value of the orbit integral by a factor of two,
provided that the coupling function has sufficiently non-pathological
behavior (such as that given by equation [B8]).

\vskip 1.0truein
\centerline{\bf REFERENCES}
\medskip

\par\pp
Adams, F. C. 1990, {\sl ApJ}, {\bf 363}, 578

\par\pp
Adams, F. C., Galli, D., Najita, J., Lizano, S., \& Shu, F. H. 1995,
{\sl ApJ}, in preparation

\par\pp
Adams, F. C., Lada, C. J., \& Shu, F. H. 1987, {\sl ApJ},
{\bf 312}, 788 (ALS)

\par\pp
Adams, F. C., Ruden, S. P., \& Shu, F. H. 1989, {\sl ApJ}, {\bf 347}, 959

\par\pp
Adams, F. C., \& Shu, F. H. 1985, {\sl ApJ}, {\bf 296}, 655

\par\pp
Adams, F. C., \& Shu, F. H. 1986, {\sl ApJ}, {\bf 308}, 836

\par\pp
Andr{\' e}, P., \& Montmerle, T. 1994, {\sl ApJ}, {\bf 420}, 837

\par\pp
Allen, C. W. 1976, Astrophysical Quantities (London: Athlone Press)

\par\pp
Butner, H. M., Evans, N. J., Lester, D. F., Levreault, R. M.,
\& Strom, S. E. 1991, {\sl ApJ}, {\bf 376}, 676

\par\pp
Cassen, P., \& Moosman, A. 1981, {\sl Icarus}, {\bf 48}, 353

\par\pp
Chevalier, R. 1983, {\sl ApJ}, {\bf 268}, 753

\par\pp
Churchwell, E., Wolfire, M. G., \& Wood, D.O.S. 1990,
{\sl ApJ}, {\bf 354}, 247

\par\pp
Draine, B. T., \& Lee, H. M. 1984, {\sl ApJ}, {\bf 285}, 89

\par\pp
Ezer, D., \& Cameron, A.G.W. 1967, {\sl Canadian J. Phys.},
{\bf 45}, 3429

\par\pp
Galli, D., \& Shu, F. H. 1993a, {\sl ApJ}, {\bf 417}, 220

\par\pp
Galli, D., \& Shu, F. H. 1993b, {\sl ApJ}, {\bf 417}, 243

\par\pp
Ho, P. T., Terebey, S., \& Turner 1994, {\sl ApJ}, {\bf 423}, 320

\par\pp
Hollenbach, D., Johnstone, D., Lizano, S., \& Shu, F. 1994,
{\sl ApJ}, {\bf 428}, 654

\par\pp
Hunter, C. 1977, {\sl ApJ}, {\bf 218}, 834

\par\pp
Kahn, F. D. 1974, {\sl A \& A}, {\bf 37}, 149

\par\pp
Kenyon, S. J., Calvet, N., \& Hartmann, L. 1993,
{\sl ApJ}, {\bf 414}, 676

\par\pp
Larson, R. B. 1969a, {\sl MNRAS}, {\bf 145}, 271

\par\pp
Larson, R. B. 1969b, {\sl MNRAS}, {\bf 145}, 297

\par\pp
Larson, R. B., \& Starrfield, S. 1971, {\sl A \& A}, {\bf 13}, 190

\par\pp
Larson, R. B. 1985, {\sl MNRAS}, {\bf 214}, 379

\par\pp
Lizano, S., \& Shu, F. H. 1989, {\sl ApJ}, {\bf 342}, 834

\par\pp{}
Mouschovias, T. Ch. 1978, in Protostars and Planets,
ed. T. Gehrels (Tucson: University of Arizona Press), p. 209

\par\pp
Myers, P. C., \& Fuller, G. A. 1992, {\sl ApJ}, {\bf 396}, 631

\par\pp
Nakano, T. 1984, {\sl Fund. Cosmic Phys.}, {\bf 9}, 139

\par\pp
Nakano, T. 1989, {\sl ApJ}, {\bf 345}, 464

\par\pp
Nakano, T., Hasegawa, T., \& Norman, C. 1995, {\sl ApJ}, in press

\par\pp
Palla, F., \& Stahler, S. W. 1990, {\sl ApJ}, {\bf 360}, L47

\par\pp
Phillips, A. C. 1994, The Physics of Stars (Chichester: Wiley)

\par\pp
Reipurth, B., Chini, R., Kr{\" u}gel, E., Kreysa, E., \& Sievers, A.
1993, {\sl A \& A}, {\bf 273}, 221

\par\pp
Shu, F. H. 1977, {\sl ApJ}, {\bf 214}, 488

\par\pp
Shu, F. H. 1983, {\sl ApJ}, {\bf 273}, 202

\par\pp
Shu, F. H., Adams, F. C., \& Lizano, S. 1987, {\sl A R A \& A},
{\bf 25}, 23

\par\pp
Shu, F. H., Tremaine, S., Adams, F. C., \& Ruden, S. P. 1990,
{\sl ApJ}, {\bf 358}, 495

\par\pp
Stahler, S. W., Shu, F. H., \& Taam R. E. 1980, {\sl ApJ}, {\bf 241}, 637

\par\pp
Terebey, S., Chandler, C. J., \& Andr{\' e}, P. 1993,
{\sl ApJ}, {\bf 414}, 759

\par\pp
Terebey, S., Shu, F. H., \& Cassen, P. 1984, {\sl ApJ}, {\bf 286}, 529

\par\pp
Ulrich, R. K. 1976, {\sl ApJ}, {\bf 210}, 377

\par\pp
Wolfire, M. G., \& Cassinelli, J. 1986, {\sl ApJ}, {\bf 310}, 207

\par\pp
Wolfire, M. G., \& Cassinelli, J. 1987, {\sl ApJ}, {\bf 319}, 850

\par\pp
Wolfire, M. G., \& Churchwell, E. 1994, {\sl ApJ}, {\bf 427}, 889

\par\pp
Yorke, H. W. 1979, {\sl A \& A}, {\bf 80}, 308

\par\pp
Yorke, H. W., \& Kr{\"u}gel, E. 1977, {\sl A \& A}, {\bf 54}, 183

\par\pp
Zuckerman, B., \& Palmer, P. 1974, {\sl A R A \& A}, {\bf 12}, 279

%\newpage
\vskip 1.0truein
\centerline{\bf FIGURE CAPTIONS}
\bigskip

\medskip
\noindent
Figure 1.  Effects of radiation pressure on an infalling trajectory
in the meridional plane.  Solid curves show the projected orbits
for a given initial angle $\theta_0$ and for varying values of
radiation pressure; shown here are curves for $\alpha^2 / R_C$ =
0.25, 0.50, and 1.0.  The dotted curve shows the orbit for the
case of no radiation pressure $\alpha = 0$.  The star is located
at the origin of the coordinates.
(a) Initial angle $\theta_0$ = $\pi/4$ ($\mu_0 = \sqrt{2}/2$).
(b) Initial angle $\theta_0$ = $\pi/8$.

\medskip
\noindent
Figure 2. The ``turn around surface'' for varying amounts of
radiation pressure.  The dotted curves show the surfaces at which
orbits are reversed due to radiation pressure.  The different
curves correspond to varying amounts of radiation pressure given
by $\alpha^2/R_C$ = 0.25 -- 2.0 in increments of 0.25.
The solid curves show representative trajectories (orbits)
with initial angles $\theta_0$ in increments of $\pi/16$
(for the case with $\alpha^2/R_C$ = 0.75).

\medskip
\noindent
Figure 3. Density profiles including radiation pressure.
The various curves show the density as a function of radius for
angles $\theta$ = 0, $\pi/6$, $\pi/4$, and $\pi/3$.
The horizontal axis shows the radius in units of the centrifugal
radius $R_C$; the vertical axis shows the density scaled to the
value $\rho_C$, which is the density distribution from the
spherical solution evaluated at the centrifugal radius
($\rho_C = C R_C^{-3/2}$).
(a) Density distribution for $\alpha^2/R_C$ = 0.
(b) Density distribution for $\alpha^2/R_C$ = 0.5.
(c) Density distribution for $\alpha^2/R_C$ = 1.0.

\medskip
\noindent
Figure 4. The asphericity function ${\cal A} (r)$ for varying
amounts of radiation pressure.  The function ${\cal A}$ is the
ratio of the spherically averaged density profile $\rhobar$ to the
density profile obtained for purely spherical collapse, i.e.,
${\cal A} = \rhobar \, r^{3/2} \, C^{-1}$.  The radius (plotted on the
horizontal axis) is given in units of the centrifugal radius $R_C$.
The different curves are for varying amounts of radiation pressure
with $\alpha^2/R_C$ = 0, 0.1, 0.3, and 0.5.

\medskip
\noindent
Figure 5. Maximum mass of a forming star for purely isothermal
collapse.  This figure shows the plane of initial conditions,
i.e., the sound speed $a$ constitutes the horizontal axis and
the rotation rate $\Omega$ constitutes the vertical axis.
The labeled curves show the maximum mass of a star that can
form with the given values of the initial conditions.  The
region in the upper left part of the diagram
has no upper mass limit.

\medskip
\noindent
Figure 6. Maximum mass of a forming star for purely logatropic
collapse.  This figure shows the plane of initial conditions,
i.e., the pressure scale $P_0$ constitutes the horizontal axis and
the rotation rate $\Omega$ constitutes the vertical axis.
The labeled curves show the maximum mass of a star that can
form with the given values of the initial conditions.  The
region in the upper left part of the diagram
has no upper mass limit.

\medskip
\noindent
Figure 7. Maximum mass of a forming star for an isothermal collapse
flow which includes radiation pressure and magnetic fields.  This
figure shows the plane of initial conditions, i.e., the sound speed
$a$ constitutes the horizontal axis and magnetic field strength $B$
constitutes the vertical axis.  The labeled curves show the maximum
mass of a star that can form with the given values of the initial
conditions.  The region in the upper left part of the diagram
has no upper mass limit.

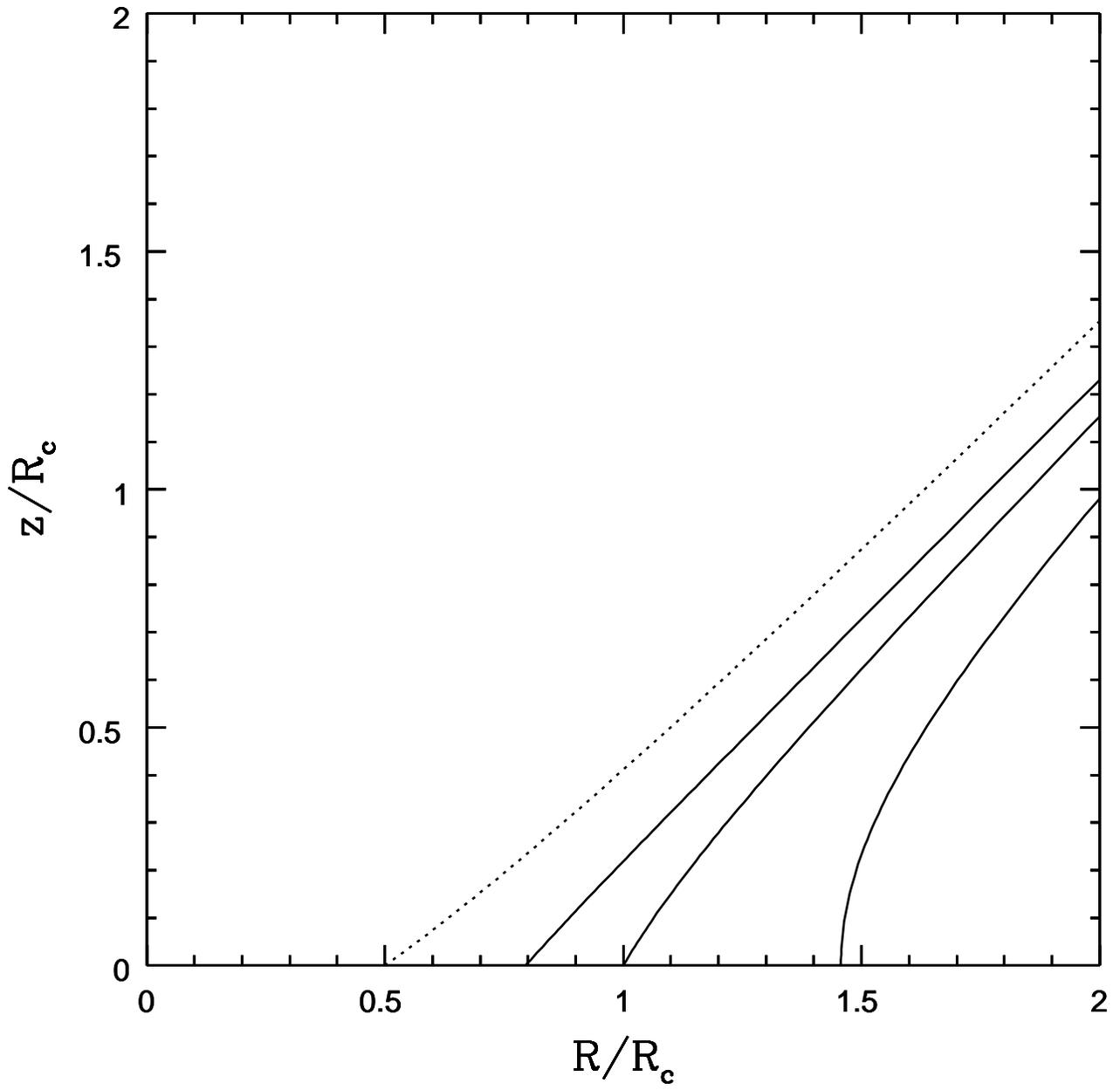
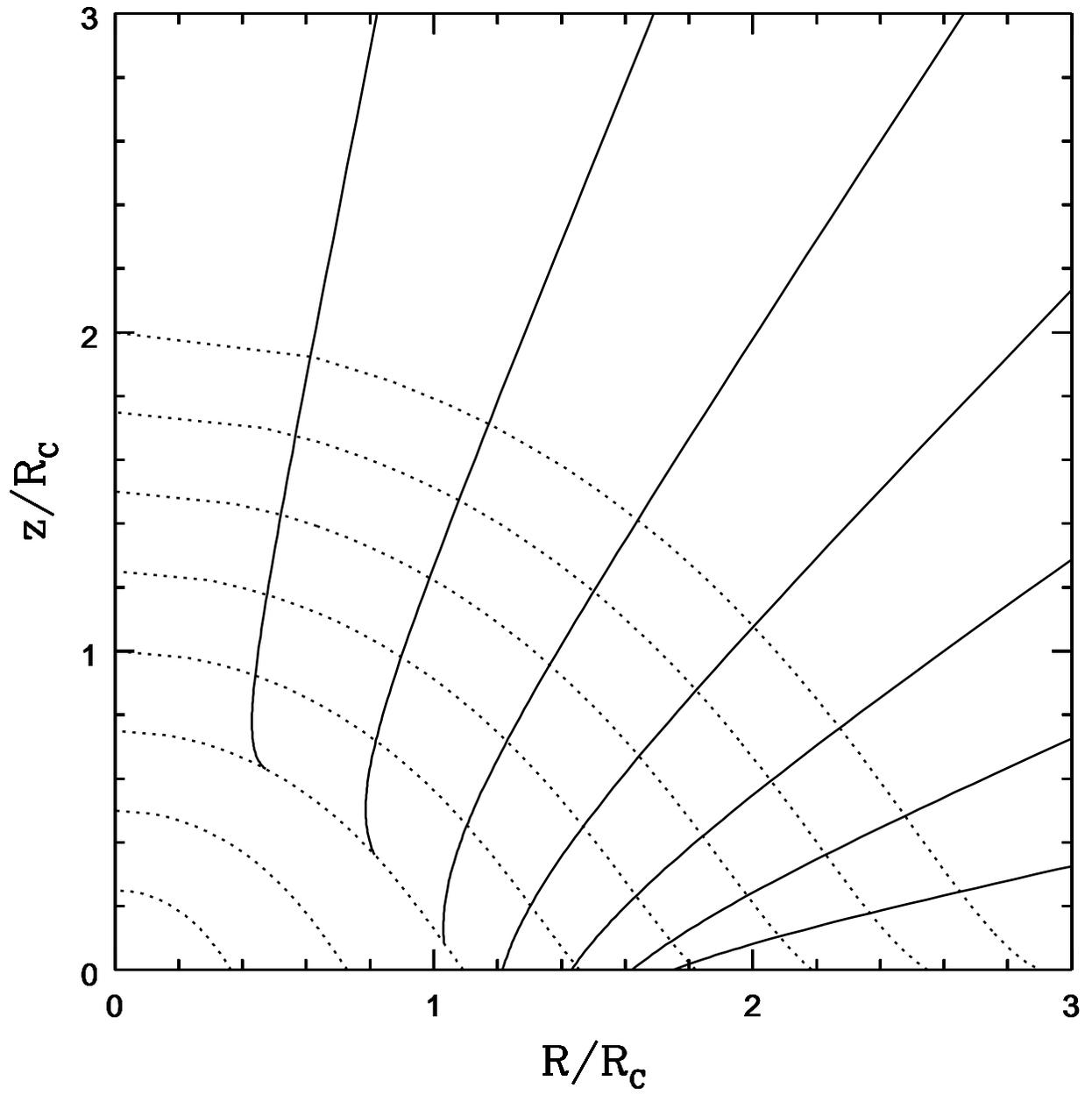
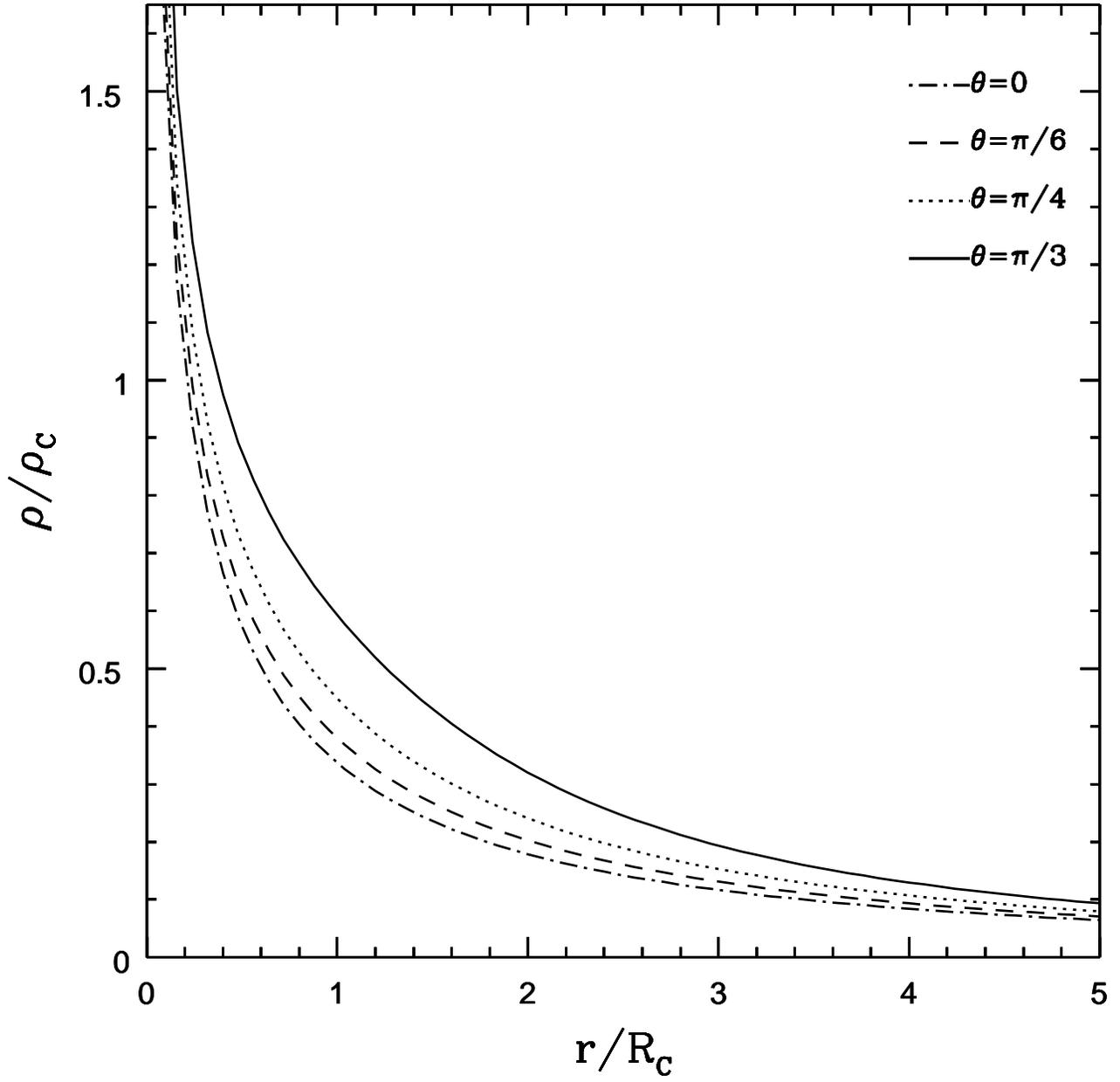
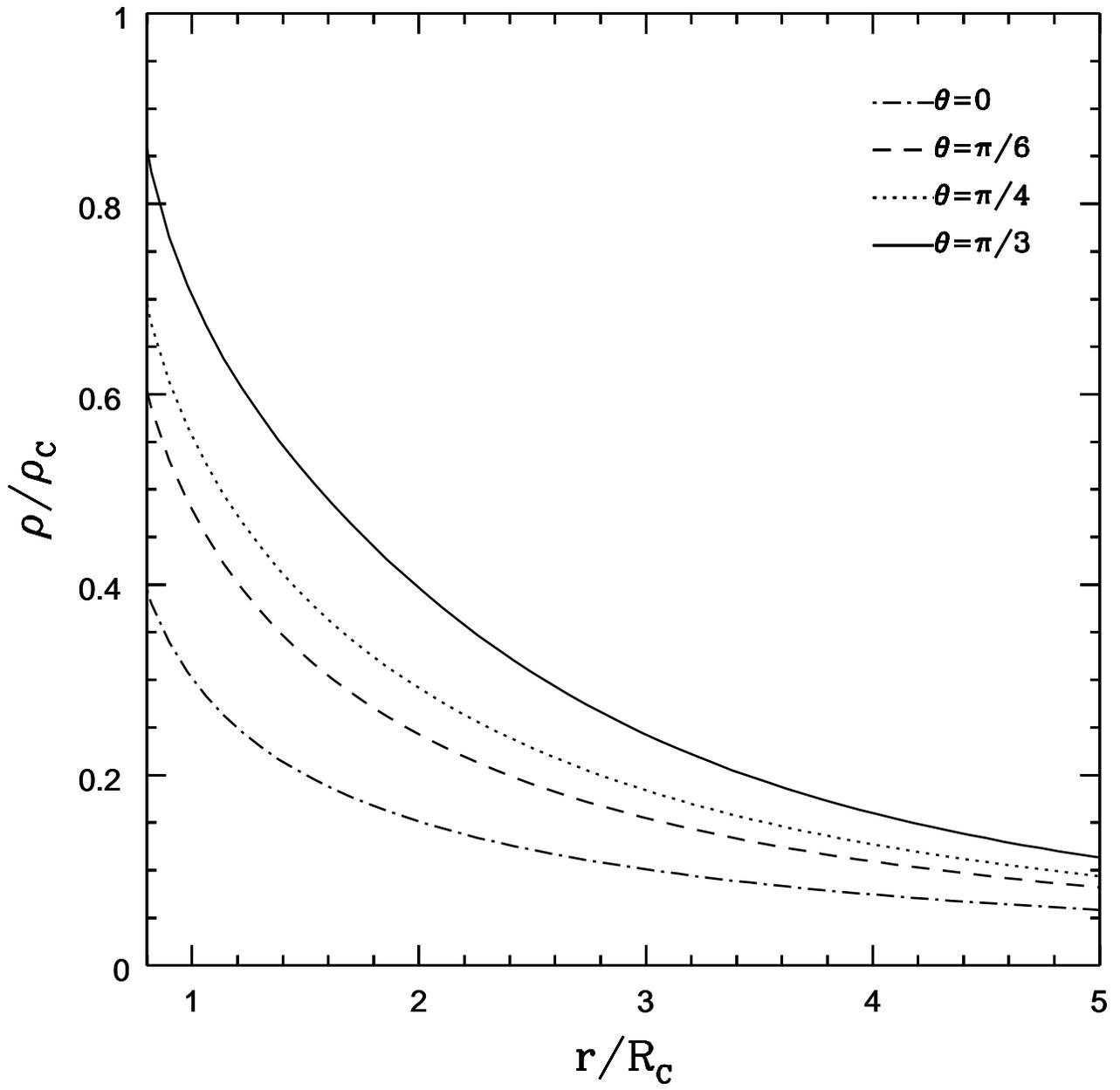
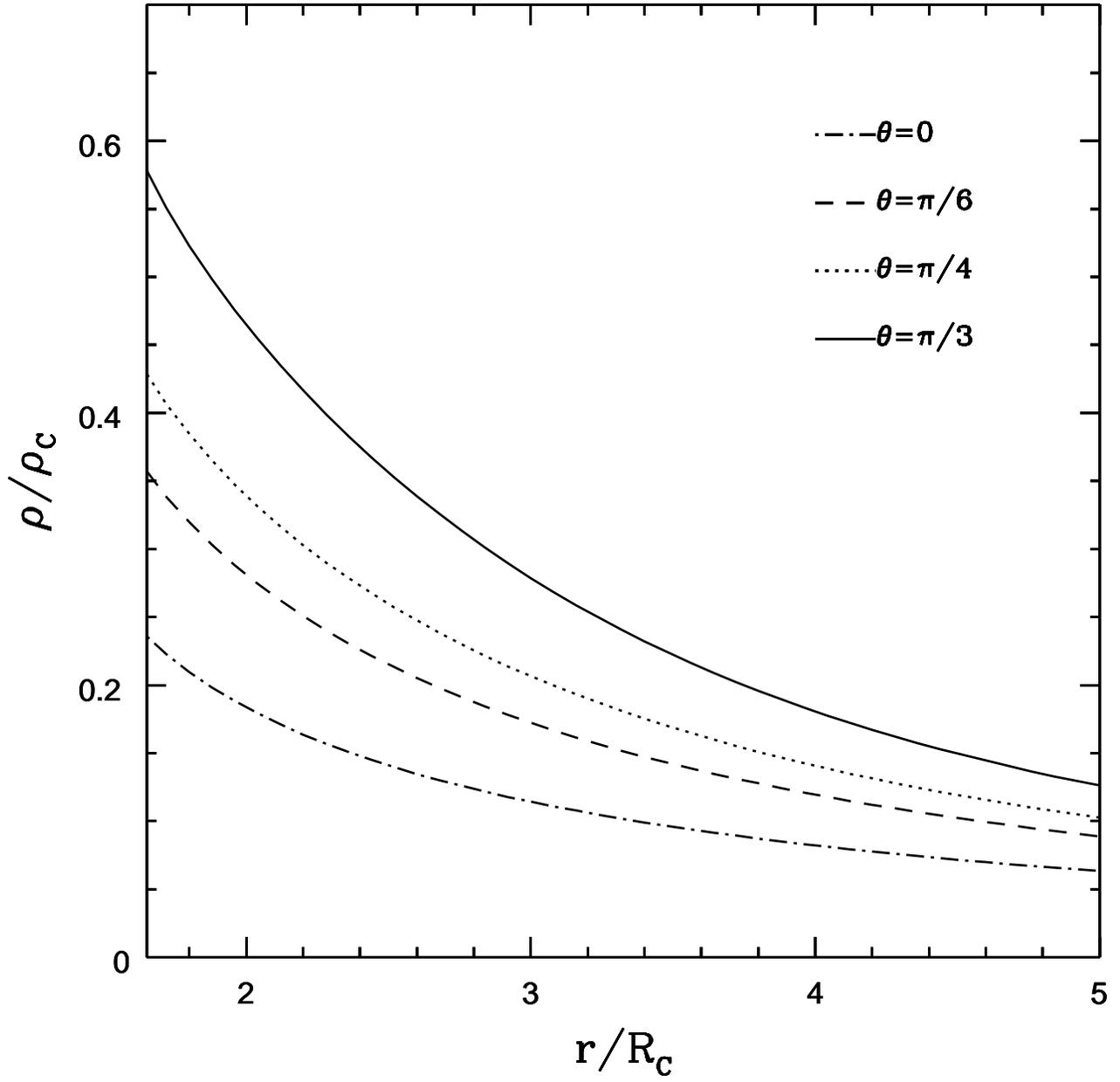
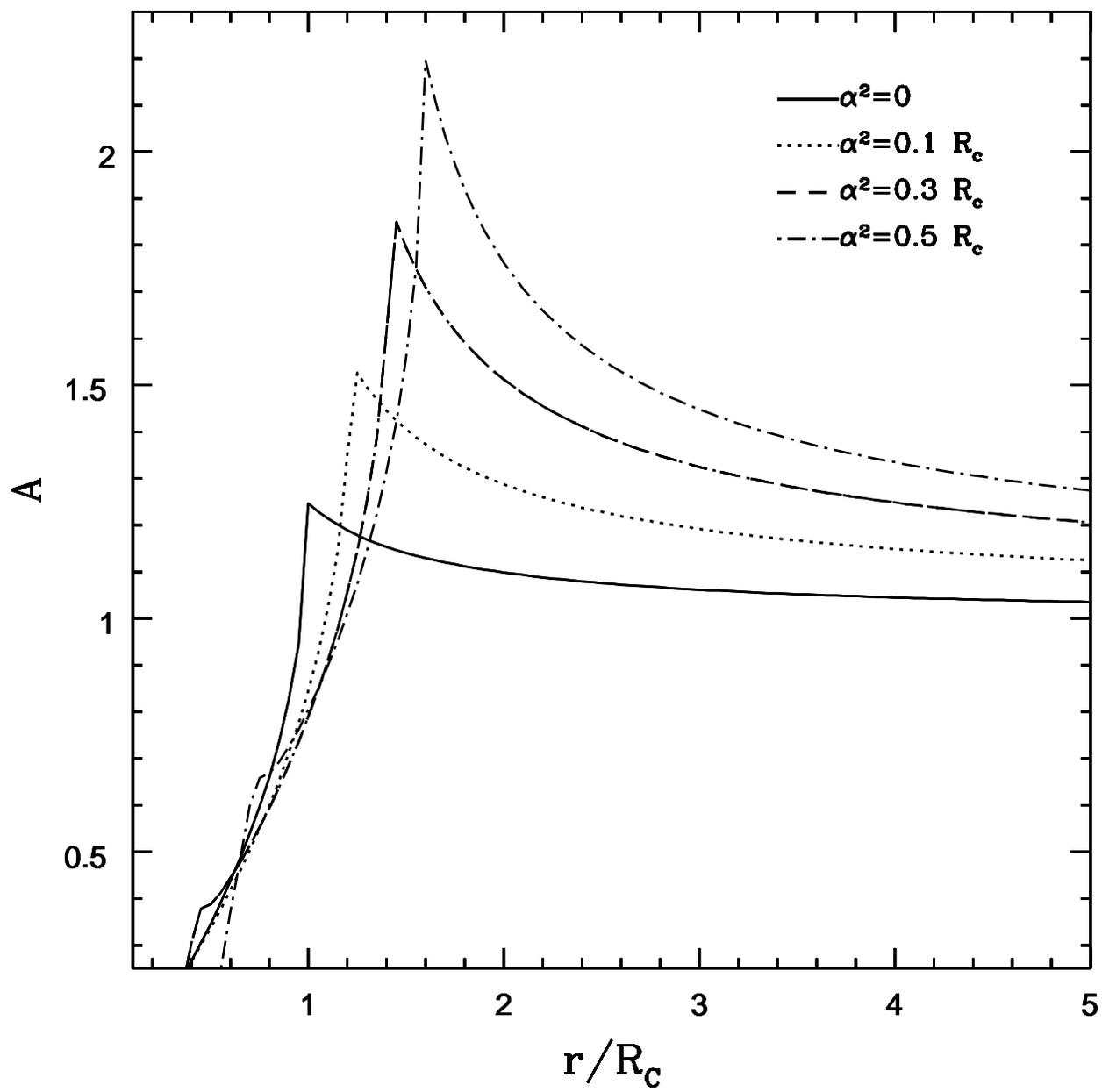
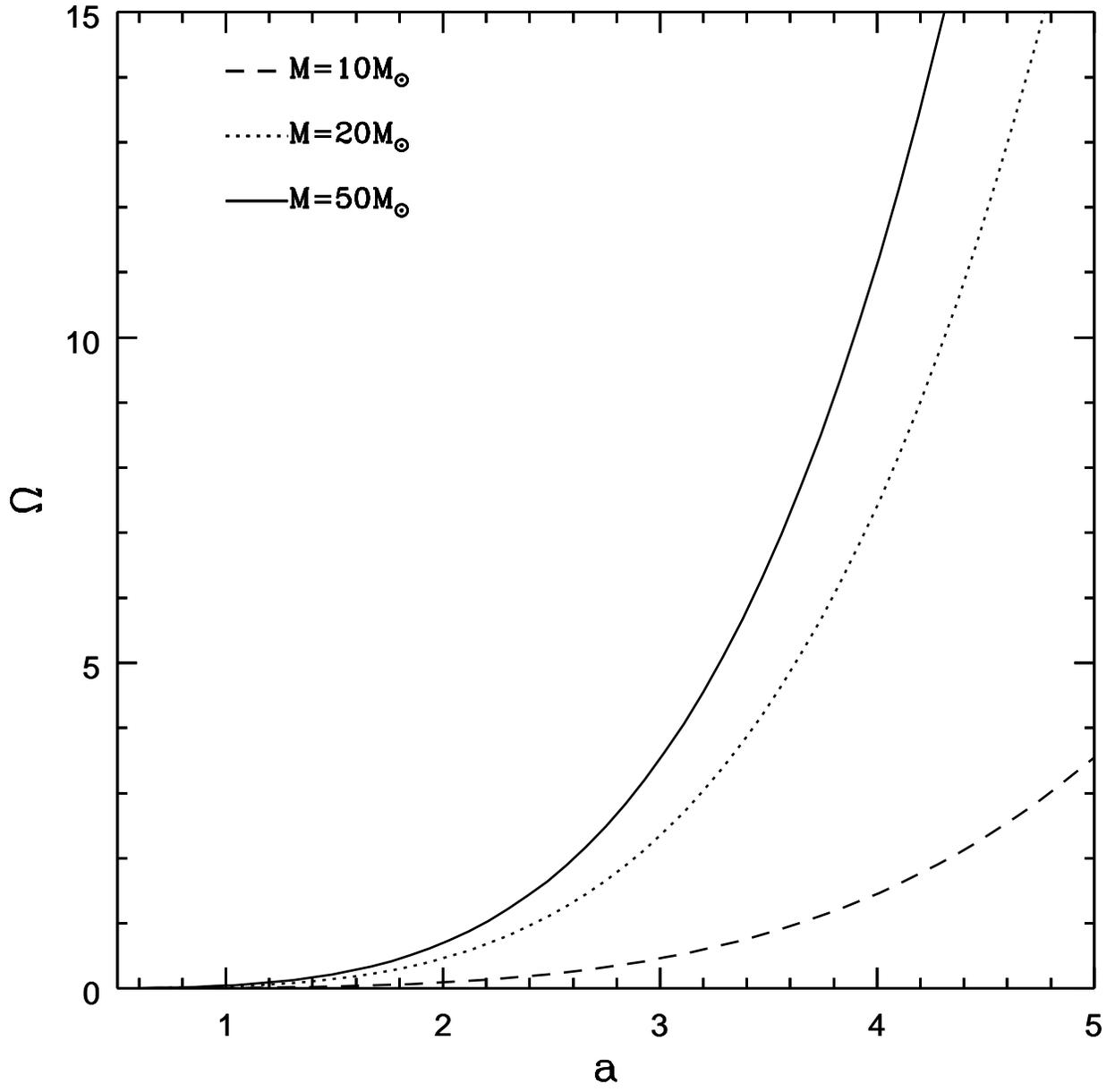
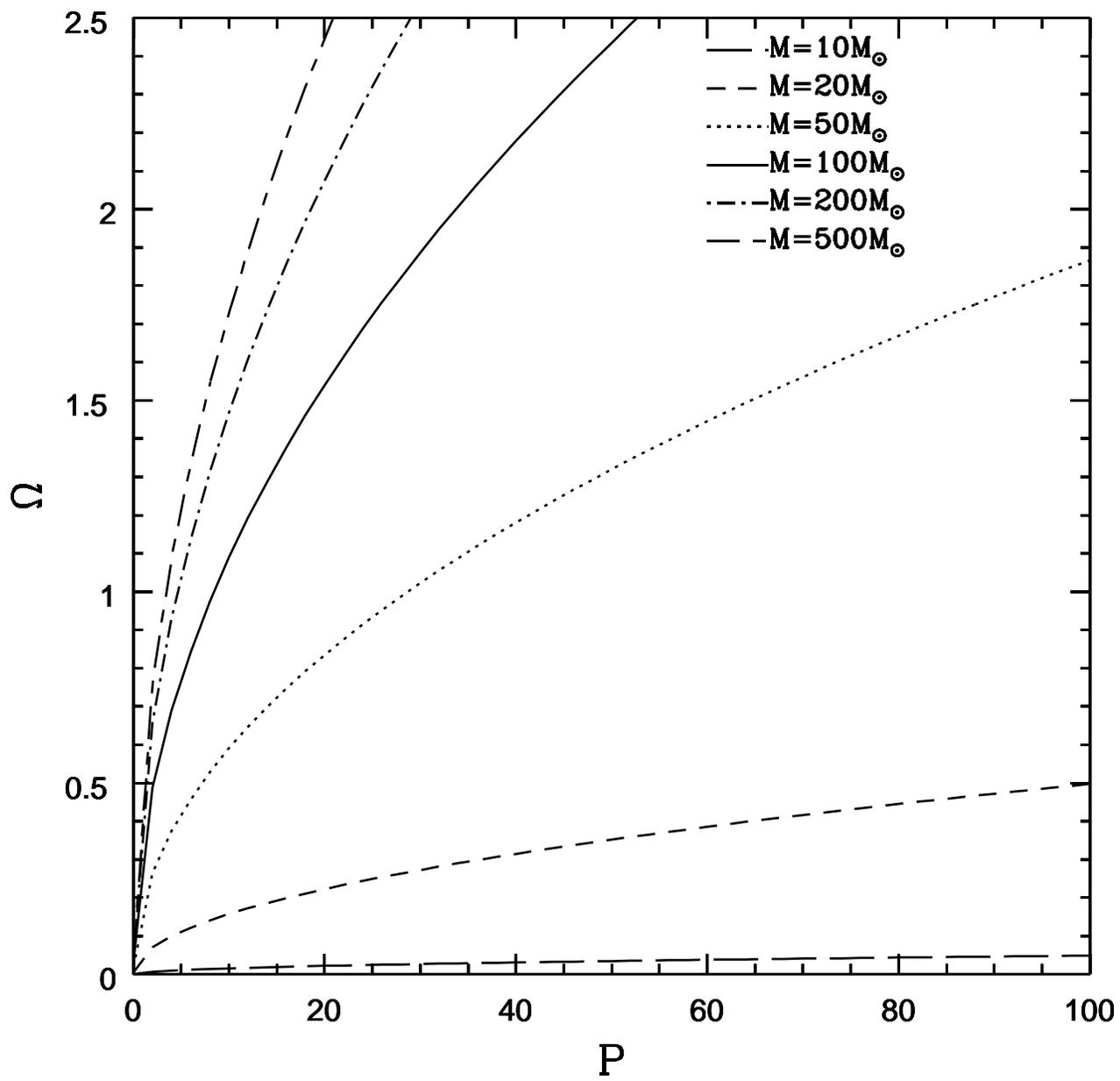
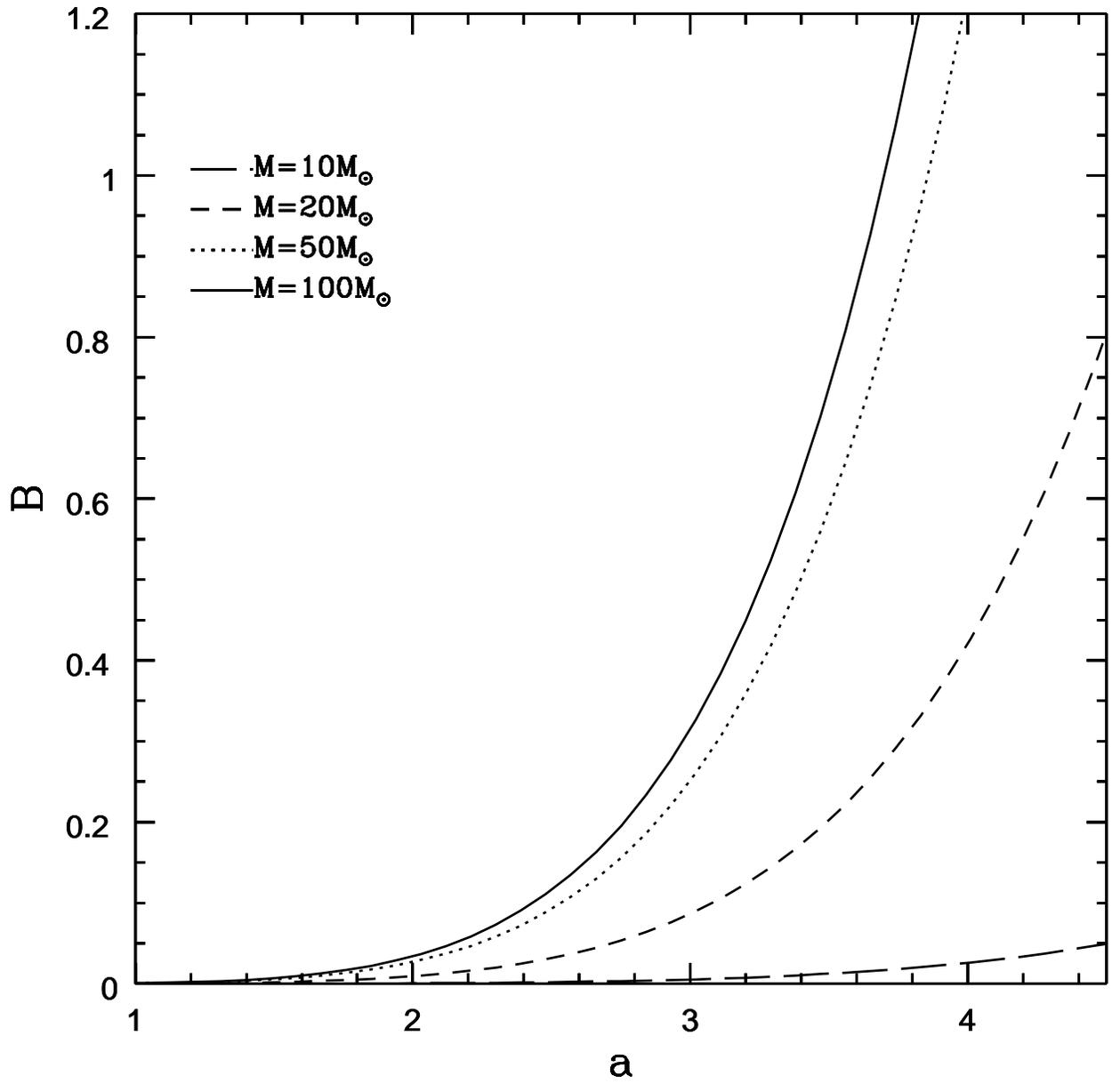
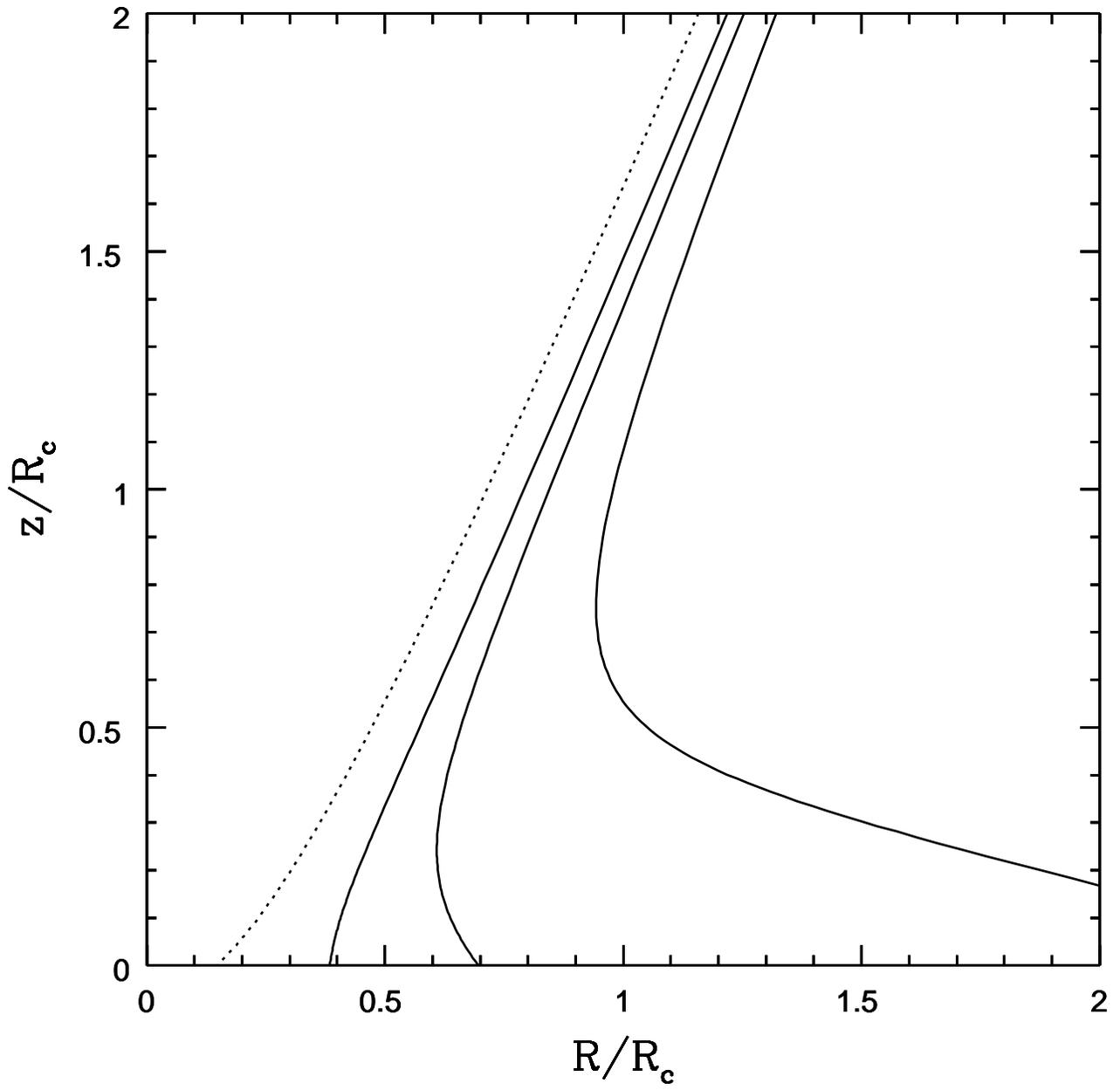
\bye